\documentclass[amsmath,amssymb,superscriptaddress,showpacs, aps]{revtex4-1} %notitlepage]

\usepackage[a4paper,hmargin={1.5cm,1.5cm},vmargin={1.5cm,2.5cm}]{geometry}
\usepackage{hyperref}
\usepackage{graphicx}
\usepackage{bm}
\newcommand{\nth}{n_{\mathrm{th}}}
\newcommand{\be}{\begin{equation}}
\newcommand{\ee}{\end{equation}}

\begin{document}

\title{Ultrastrong coupling of a single artificial atom to an electromagnetic continuum in the nonperturbative regime}

\author{P.~Forn-D\'iaz}
\address{Institute for Quantum Computing, Department of Physics and Astronomy, and Waterloo Institute for Nanotechnology, University of Waterloo, Waterloo, N2L 3G1, Canada}
\author{J.~J.~Garc\'ia-Ripoll}
\address{Instituto de F\'isica Fundamental IFF-CSIC, Madrid 28006, Spain}
\author{B.~Peropadre}
\address{Department of Chemistry and Chemical Biology, Harvard University, Cambridge, Massachusetts 02138, United States}
\author{J.-L.~Orgiazzi}
\address{Institute for Quantum Computing and Department of Electrical and Computer Engineering, University of Waterloo, Waterloo, N2L 3G1, Canada}
\author{M.~A.~Yurtalan}
\address{Institute for Quantum Computing and Department of Electrical and Computer Engineering, University of Waterloo, Waterloo, N2L 3G1, Canada}
\author{R.~Belyansky}
\address{Institute for Quantum Computing and Department of Electrical and Computer Engineering, University of Waterloo, Waterloo, N2L 3G1, Canada}
\author{C.~M.~Wilson}
\thanks{These authors contributed equally to this work.}
\address{Institute for Quantum Computing and Department of Electrical and Computer Engineering, University of Waterloo, Waterloo, N2L 3G1, Canada}
\author{A.~Lupascu}
\thanks{These authors contributed equally to this work.}
\address{Institute for Quantum Computing, Department of Physics and Astronomy, and Waterloo Institute for Nanotechnology, University of Waterloo, Waterloo, N2L 3G1, Canada}

\begin{abstract}
The study of light-matter interaction has led to many fundamental discoveries as well as numerous important technologies. Over the last decades, great strides have been made in increasing the strength of this interaction at the single-photon level, leading to a continual exploration of new physics and applications. Recently, a major achievement has been the demonstration of the so-called strong coupling regime \cite{paris_book,wallraff}, a key advancement enabling great progress in quantum information science. Here, we demonstrate light-matter interaction over an order of magnitude stronger than previously reported, reaching the nonperturbative regime of ultrastrong coupling (USC). We achieve this using a superconducting artificial atom tunably coupled to the electromagnetic continuum of a one-dimensional waveguide. For the largest coupling, the spontaneous emission rate of the atom exceeds its transition frequency. In this USC regime, the description of atom and light as distinct entities breaks down, and a new description in terms of hybrid states is required \cite{borja_usc,juanjo_polaron}. Our results open the door to a wealth of new physics and applications. Beyond light-matter interaction itself, the tunability of our system makes it a promising tool to study a number of important physical systems such as the well-known spin-boson \cite{spin-boson_rmp} and Kondo models \cite{kondo}.
\end{abstract}

\maketitle

Light propagating in a one-dimensional (1D) waveguide is described by a 1D electromagnetic field with a continuous spectrum of frequencies. The strong coupling regime \cite{shen_fan} between an atom and such an electromagnetic continuum is defined as the regime in which the atom emits radiation predominantly into the waveguide with a rate $\Gamma_G$ that significantly exceeds the decoherence rate of the atom as well as emission into any other channel. In this regime, the atomic transition frequency $\Delta$ far exceeds the emission rate $\Gamma_G\ll\Delta$. Achieving strong coupling to a continuum is a recent achievement in quantum optics \cite{astafiev_fluor}. Strong atom-waveguide coupling has numerous applications such as the development of quantum networks \cite{kimble_qint} for quantum communication \cite{hoi_router} and quantum simulation \cite{laura}. This technology, first demonstrated with superconducting qubits in open transmission lines \cite{astafiev_fluor, hoi_router, hoi_kerr, arjan_2tr}, has also been implemented with both neutral atoms \cite{aki_super}, and quantum dots \cite{lodahl_qdot} in photonic crystal waveguides. The distinctive signature of strong coupling is a decrease below 50\% of the amplitude of transmitted signals due to coherent atomic scattering of photons. 

A distinct regime of light-matter interaction is reached when $\Gamma_G$ becomes comparable to the atomic transition frequency $\Gamma_G/\Delta\sim0.1$, the ultrastrong coupling (USC) regime. Most studies involving atom-field interactions are in the regime $\Gamma_G\ll\Delta$ where the common rotating-wave approximation (RWA) applies. In the USC regime, the RWA breaks down but perturbative treatments still allow an effective atom-field description when $\Gamma_G/\Delta\sim0.1$ \cite{niem, bloch-siegert}. A novel, unexplored regime of light-matter interaction is the nonperturbative USC regime, where $\Gamma_G$ approaches or exceeds the atomic transition frequency $\Gamma_G/\Delta\sim1$ and perturbation theory breaks down. This is a general definition also applicable to the case of discrete modes in cavity-QED systems \cite{ciuti_prb}. We note that the nonperturbative USC regime has also been referred to in the literature as the deep strong coupling regime \cite{jorge_dsc}. In the nonperturbative USC regime, the atom-photon system is described by photons dressing the atom even in the ground state \cite{ciuti_prb, borja_usc, juanjo_polaron}. In this regime, the Markovian approximation also breaks down because the broad qubit linewidth $\Gamma_G$ implies that the spectral density of the environment seen by the atom is not independent of frequency. 
The presence of a continuum of modes ultrastrongly coupled to an atom has the additional effect of renormalizing the atomic frequency from the bare value $\Delta_0$, which is a generalization of the well-known Lamb shift to arbitrary coupling strengths. 
These renormalization effects are also central to the well-known spin-boson model \cite{spin-boson_rmp}, which has been used to describe, for example, open quantum systems \cite{grifoni_SB}, quantum stochastic resonance \cite{rmp_hanggi} and phase transitions in Josephson junctions \cite{penttila}. Reaching the nonperturbative USC regime allows the exploration of the ultimate limits in light-matter interaction strength and relativistic quantum information phenomena \cite{edu}. 
In addition, ultrastrong couplings may have technological applications, such as single-photon nonlinearities \cite{sanchez_burillo} and broadband single-photon sources \cite{borja_usc}.  

Superconducting qubits are artificial atoms with transitions in the microwave range of frequencies. Recently, flux-type superconducting qubits have been put forward as candidates to reach the nonperturbative USC regime \cite{bourassa, borja_usc}, having demonstrated large galvanic couplings to resonators \cite{bloch-siegert} and a large anharmonicity that allows them to remain an effective two-level system when $\Gamma_G\sim\Delta$. This is in contrast to other more weakly anharmonic qubits whose transitions would overlap for large enough $\Gamma_G$. 

\begin{figure}[!hbt]
\centering
\includegraphics{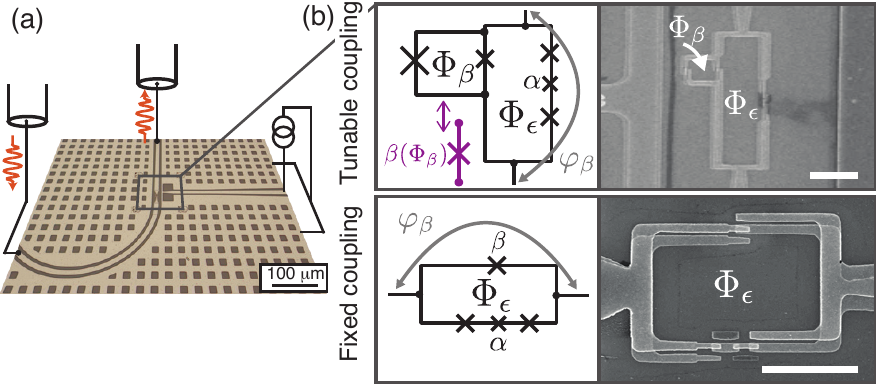}
\caption{\label{fig1}Measurement setup and devices. (a) Schematic of the circuit layout, with a micrograph of a section of a chip containing a transmission line and a flux qubit. (b) Circuit schematic of a flux qubit coupled to a transmission line with tunable (fixed) coupling shown at the top (bottom). In both cases, the coupling is proportional to the matrix element of the phase operator $\varphi_{\beta}$ across the coupling junction $\beta$. The scanning electron micrographs show the corresponding circuits. The white scale bars are 4~$\mu$m.}  
\end{figure} 
Here, we demonstrate nonperturbative ultrastrong coupling of a superconducting flux qubit \cite{hans_sci} coupled to an open 1D transmission line via a shared Josephson junction. As predicted \cite{bourassa, borja_usc}, we observe that $\Gamma_G$ scales with the inverse of the coupling junction size. For devices with a small-enough coupling junction we measure $\Gamma_G\sim\Delta$, indicating that we reach the nonperturbative USC regime. Our flux qubit has four Josephson junctions. Two reference junctions are designed with the same area, while the areas of the other two are scaled by the factors $\alpha\sim0.6$ and $\beta>1$ with respect to the area of the reference junctions \cite{JL_dec}. The flux qubit is galvanically attached to the center line of a 1D coplanar waveguide transmission line (Fig.~1(a)). In order to achieve ultrastrong couplings, we place the $\beta$-junction in parallel to the other three (Fig.~1(b)). The coupling to the line is then
mainly determined \cite{bourassa, borja_usc} by the matrix element between ground $|0\rangle$ and excited $|1\rangle$ qubit states of the superconducting phase operator across the $\beta$-junction $\langle0|\hat{\varphi}_{\beta}|1\rangle\equiv\varphi_{\beta}$, which is the dominant contribution to the coupling for $\beta<4$. 
Further, we make the coupling tuneable by turning the $\beta$-junction into a superconducting quantum interference device (SQUID) threaded by a flux $\Phi_{\beta}$ as shown in Fig.~1(b) (Methods). 

The experiments are performed by applying a probe field with a variable frequency and recording the transmitted field amplitude and phase on a vector network analyzer. For emission rates $\Gamma_1/\Delta\ll1$, where $\Gamma_1$ is the total emission rate, and in the presence of thermal excitations, the transmitted coherent scattering amplitude at low driving power is given by \cite{astafiev_fluor, borja_scatt}:
\begin{equation}\label{tr_scatt}
T  = 1+R \approx
\frac{1+(\delta\omega/\Gamma_2)^2+r_0(i\delta\omega/\Gamma_2-1)}{1+(\delta\omega/\Gamma_2)^2}.
\end{equation}
Here $\Gamma_2\equiv\Gamma_{\mathrm{\varphi}}+(\Gamma_1/2)(1+2n_{\mathrm{th}})$ is the total decoherence rate, $\Gamma_{\mathrm{\varphi}}$ is the pure dephasing rate, $\delta\omega=\omega-\Delta$ is the detuning of the probe field, and $n_{\mathrm{th}}$ is the thermal photon occupation number at the qubit frequency (Supplementary Information). The maximum reflection amplitude is $r_0=\Gamma_1/[2\Gamma_2(1+2n_{\rm{th}})]$. As in other experiments on superconducting quantum circuits \cite{astafiev_fluor, hoi_router}, relaxation into channels other than the waveguide is negligible. Therefore, we assume $\Gamma_1 = \Gamma_G$. We note that equation~(1) applies in the RWA. However, it has recently been shown \cite{juanjo_polaron} that the scattering line shapes are approximately Lorentzian in the USC regime up to $\Gamma_1/\Delta\sim1$ if we consider $\Delta$ and $\Gamma_1$ to be renormalized parameters. This can be shown using a polaron transformation, allowing us to interpret the scattering center as an atom dressed by a cloud of photons. 

We first show measurements on a device with a fixed coupling junction with $\beta\simeq3.5$ (Fig.~1(b)). The transmission spectrum as a function of applied magnetic field (Fig.~2(a)) shows a maximum extinction at the symmetry point of 95\%, indicating strong coupling. By fitting equation~1 (dashed line), we infer $\Gamma_1/2\pi=88\pm11$~MHz (see Methods), $\Delta/2\pi=3.996\pm0.001~$GHz, giving $\Gamma_1/\Delta=0.02$ which is not in the USC regime. Flux qubit spectra in transmission lines similar to this one have previously been reported \cite{astafiev_fluor, garching_SB}. 
\begin{figure}[!hbt]
\centering
\includegraphics{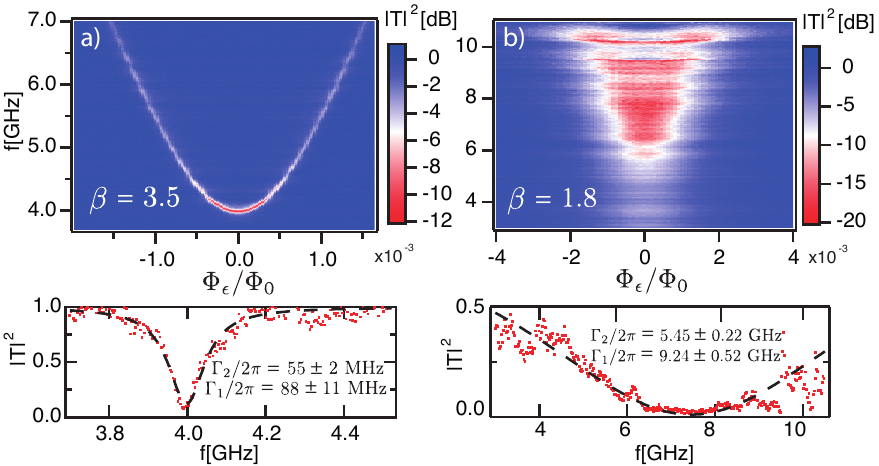}
\caption{\label{fig2}Spectroscopy of devices with fixed coupling. Top plots show transmission versus frequency and magnetic flux, referenced to $\Phi_0/2$. Bottom plots show transmission corresponding to the magnetic flux at the minimum qubit splitting. Dashed lines are fits to equation~(1). Bounds on $\Gamma_1$ are from considerations of thermal effects (Methods). (a) Transmission spectrum of qubit with $\beta\simeq3.5$ and gap $\Delta/2\pi = 3.996\pm0.001$~GHz. The 95\% extinction on-resonance indicates strong coupling. (b) Spectrum of qubit with $\beta\simeq1.8$. The fit yields $\Gamma_1/2\pi\simeq9.24\pm0.52~$GHz, exceeding the qubit gap of $\Delta/2\pi=7.68\pm0.08$~GHz. This implies $\Gamma_1/\Delta = 1.20\pm0.07$, which indicates ultrastrong coupling. The extinction of the transmitted power at the symmetry point is 97\%.}
\end{figure} 

In order to enhance the coupling strength, we designed a second device where the size of the $\beta$-junction was decreased to $\beta\simeq1.8$. The resulting qubit spectrum in Fig.~2(b) shows striking differences compared to the previous device with $\beta\simeq3.5$. The qubit linewidth at the symmetry point is very large, comparable to the total measurement bandwidth of 3-11~GHz. The deviations from a Lorentzian line shape are due to bandwidth limitations of our setup, still allowing us to infer a full width at half maximum of 2$\Gamma_2/2\pi\simeq10.90\pm0.44~$GHz (see Methods). The extracted qubit emission rate $\Gamma_1/2\pi\simeq9.24\pm0.52~$GHz exceeds the qubit splitting $\Delta/2\pi=7.68\pm0.08~$GHz, giving $\Gamma_1/\Delta=1.20\pm0.07$, a clear indication that this device reaches the nonperturbative USC regime. 
\begin{figure*}[!hbt]
\centering
\includegraphics{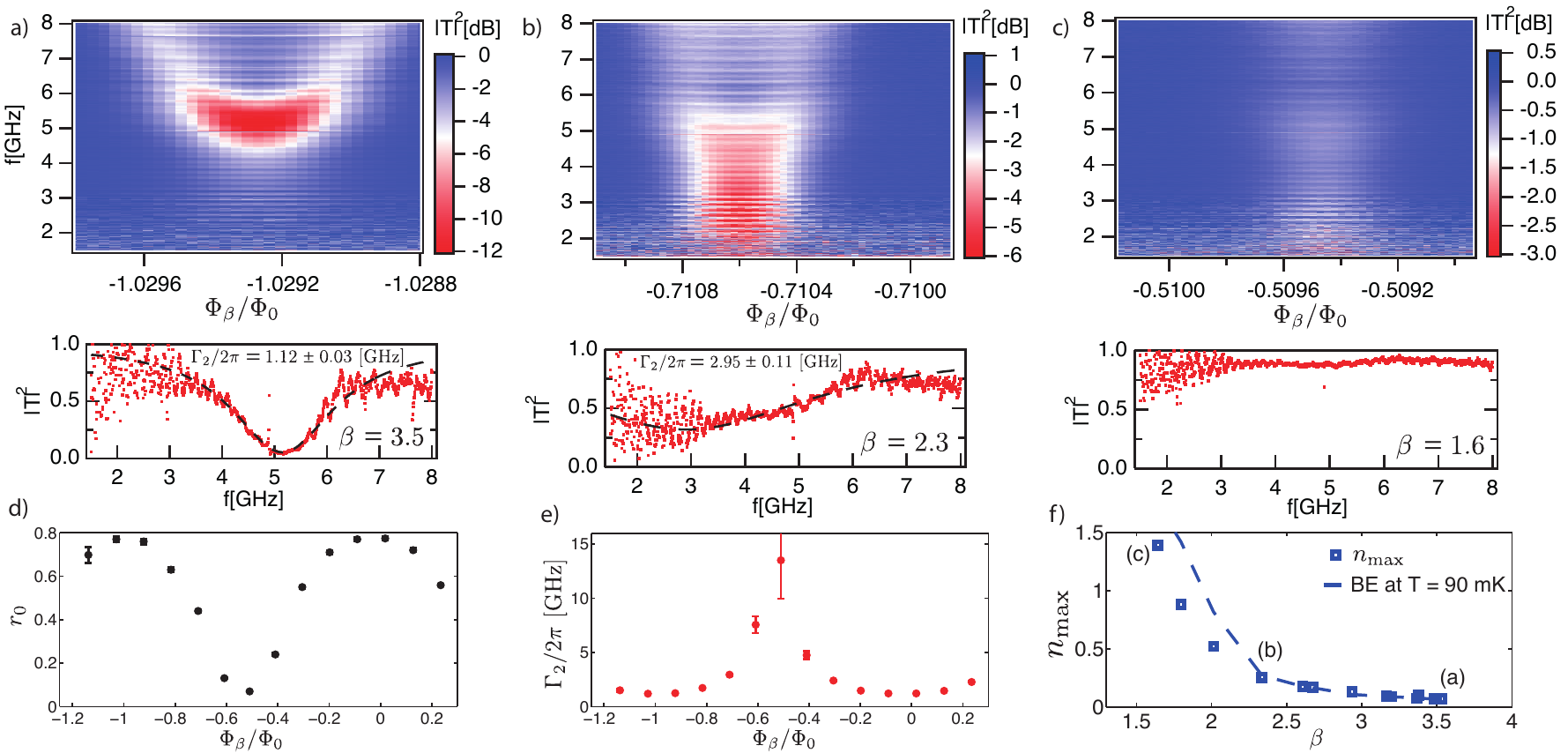}
\caption{\label{fig3}Tunable ultrastrong coupling device. (a-c) Colour plots of transmission versus frequency and magnetic flux (top) and line plots at the magnetic flux corresponding to the minimum qubit splitting (bottom). Dashed lines are fits to equation~(1). As a function of the applied magnetic field we observe a transition from strong (a) to nonperturbative ultrastrong coupling (b), (c). (a) For $\Phi_{\beta}/\Phi_0\simeq-1$ the coupling is lowest ($\beta$ largest) and the extinction is 95\% of the transmitted power. (b) At $\Phi_{\beta}/\Phi_0\simeq-0.71$ the qubit reaches $\Gamma_1\simeq\Delta.$ (c) Near $\Phi_{\beta}/\Phi_0\simeq-0.5$ the system only reflects 10\% of the incoming power and shows little signature of frequency dependence. The measured normalized couplings $\Gamma_1/\Delta$ are (a) 0.35, (b) 0.90 and (c)~$>1.5$, respectively. The large oscillations observed below 4~GHz are caused by reflections outside of our optimal measurement bandwidth 4-8~GHz. Fitting equation~(1) at the symmetry point of each qubit resonance allows extraction of the modulation of $r_0$ (d) and $\Gamma_2$ (e). Error bars represent the uncertainty in the fitted values of $r_0$ and $\Gamma_2$. From these values, we can compute bounds for $\Gamma_1$ and the maximum thermal photon number $n_{\rm{max}}$ (see Methods). (f) Extracted $n_{\rm{max}}$ showing thermal excitation at lower $\beta$ (lower frequency). Size of markers includes error bars. The decreasing value of $\Delta$ below $\sim5~$GHz causes the photon occupation to increase exponentially, closely following a Bose-Einstein (BE) distribution at $T_{\rm{eff}}=90~$mK (dash-dotted line) for $\beta>2$. The particular resonances shown in panels (a)-(c) are indicated.}
\end{figure*} 

Having observed two devices with $\Gamma_1\ll\Delta$ and $\Gamma_1>\Delta$, we now explore the intermediate region using a device with tunable coupling (Fig.~1(b)) designed with a tunable range of $\beta\sim1.6-3.6$. In Figs.~3(a)-(c), spectroscopy of the tunable coupling device is shown at three different values of $\Phi_{\beta}$. Using scanning-electron microscope (SEM) images of the measured device, we identify Figs.~3(a)-(c) as effectively having, respectively, $\beta_{\rm{(a})}\simeq3.6$, $\beta_{\rm{(b)}}\simeq2.0$, $\beta_{\rm{(c)}}\simeq1.6$. Fig.~3(a) corresponds to the highest effective $\beta$-junction size, therefore the lowest coupling strength. A flux qubit spectrum 
can be identified with $\Delta/2\pi=5.20\pm0.02$~GHz and $2\Gamma_2/2\pi\simeq2.40\pm0.07~$GHz. The maximum extinction at the symmetry point is over 95\%. The quality of the signal below 4~GHz degrades due to the measurement taking place outside the optimal bandwidth of our amplifier and circulators (4-8~GHz, Supplementary Information). In Fig.~3(b), the qubit gap decreases to $\Delta/2\pi\simeq2.90\pm0.05$~GHz as expected for a smaller $\beta$-junction. The width $2\Gamma_2/2\pi=5.90\pm0.22~$GHz is clearly enhanced, with the extinction decreasing to 30\%. In Fig.~3(c), the qubit spectrum is barely discernible. The extinction is only 10\%, with a response that appears featureless in our frequency range. Figs.~3(d),(e) show the extracted values of $r_0$ and $\Gamma_2$ using equation~1. The value of $2\Gamma_2/2\pi\simeq13\pm3~$GHz from Fig.~3(c) is an inferred bound due to the difficulty in fitting the transmission at this value of flux.

In order to understand the spectrum of the tunable coupling device and extract the corresponding emission rates $\Gamma_1$, we need to take into account finite temperature effects. We can set an upper bound on $n_{\rm{th}}$, which is $n_{\rm{max}}\equiv(1/2)(1/\sqrt{r_0}-1)$ (Methods).  Fig.~3(f) shows that the values of $n_{\rm{max}}$ for $\beta>2$ are consistent with a unique maximum effective temperature of $T_{\rm{eff}}=90$~mK, comparable to other superconducting qubit experiments. Using $0 < n_{\rm{th}} < n_{\rm{max}}$, we then put bounds on $\Gamma_1$: $2\Gamma_2r_0<\Gamma_1<2\Gamma_2\sqrt{r_0}$. Using these bounds, we plot $\Gamma_1/\Delta$ in Fig.~4(a).  The plot clearly shows that we can tune the device from the regime of strong coupling all the way into the nonperturbative USC regime. The curves in Fig.~4(a) correspond to the theoretical value of the normalized coupling strength (Supplementary Information)
\begin{equation}
\Gamma_1/\Delta\simeq\frac{1}{2\pi}\frac{R_Q}{Z_0}|\varphi_{\beta}|^2,
\end{equation}
with $R_Q=h/(2e)^2=6.5~\rm{k}\Omega$ the resistance quantum and $Z_0$ the characteristic impedance of the line. The matrix element values of the phase operator across the coupling junction $\beta$, $|\varphi_{\beta}|^2$, are calculated using the methods of reference 3. The observed values of $\Gamma_1/\Delta$ agree very well with the calculated values based on our circuit \cite{borja_usc} for an impedance close to the nominal $50~\Omega$. Above $\Gamma_1/\Delta \simeq \pi/2$, equation~(2) becomes a lower bound (Supplementary Information). This is consistent with data in the range $\beta<2$ lying above equation~(2). Including renormalization effects \cite{spin-boson_rmp} in equation~(2) might further improve the agreement with the measurements for $\beta<2$. 
\begin{figure}[!hbt]
\centering
\includegraphics{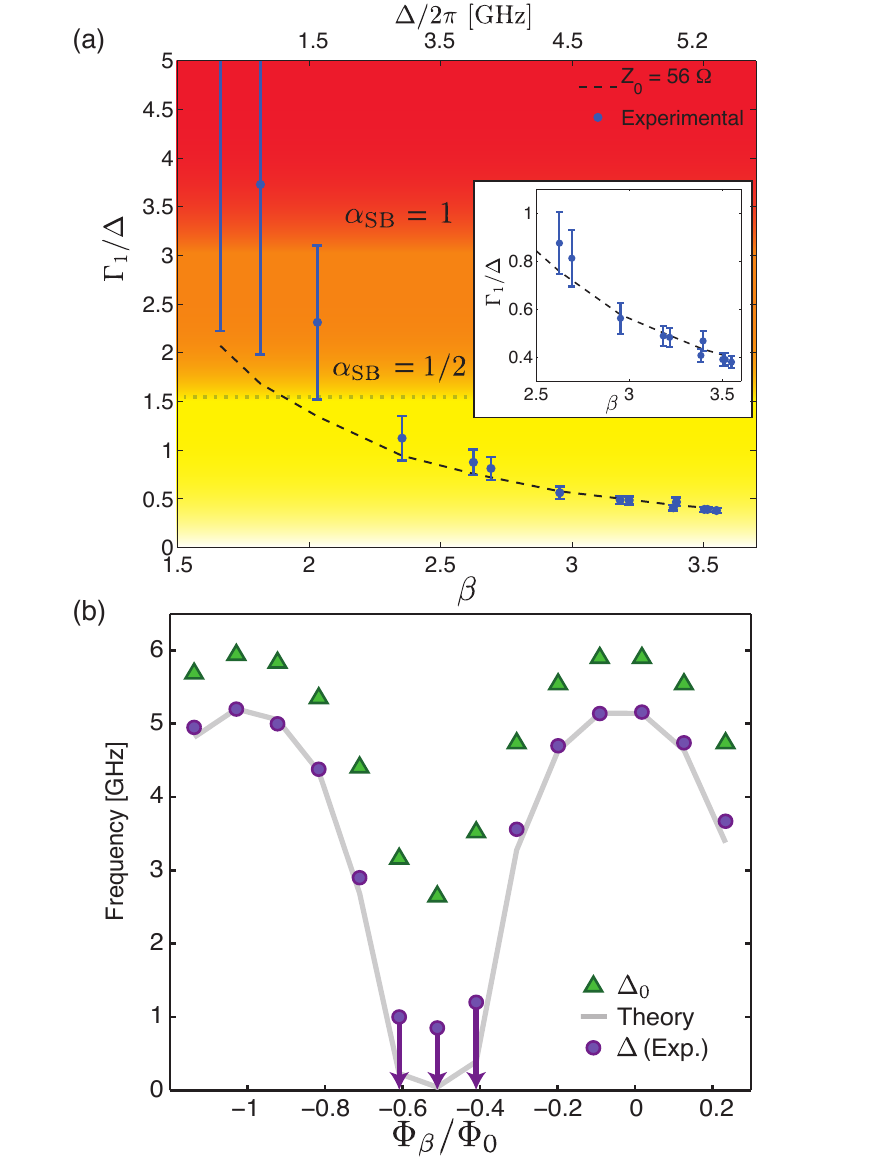}
\caption{\label{fig4}Normalized coupling rates and frequency renormalization. (a) Experimental normalized coupling rate $\Gamma_1/\Delta$ (dots) as a function of the coupling junction size $\beta$ for the device with tunable coupling. Error bars correspond to systematic bounds on $\Gamma_1$ (see Methods). The dashed curve represents the calculated parameter $\Gamma_1/\Delta$ from equation~(2). There is very good agreement with the data for an impedance close to the nominal $50~\Omega$. The colored regions indicate the spin-boson model regimes where the qubit dynamics are underdamped, overdamped and localized. The inset shows an enlargement of the high-$\beta$ region. For $\beta<2$ the curve represents a lower bound. (b) Observed qubit frequency $\Delta$ at the symmetry point (circles) as function of $\Phi_{\beta}$, along with calculated bare qubit gaps $\Delta_0$ (triangles). The curve is the theoretical prediction for the renormalized qubit gaps calculated using equation~(3) assuming a cutoff frequency of $\omega_C/2\pi=50~$GHz. Near integers of $\Phi_{\beta}/\Phi_0$, the coupling to the line is minimum and the observed $\Delta$ follows the shape of the calculated $\Delta_0$, with an offset. Near $\Phi_{\beta}/\Phi_0\sim-0.5$, the difference between $\Delta$ and $\Delta_0$ increases substantially. This is the region of nonperturbative ultrastrong coupling and the suppression of $\Delta$ is consistent with the renormalization effects predicted by the spin-boson model. The spectra in this region are difficult to fit with a Lorentzian and upper bounds to the frequency indicated by arrows are drawn instead.}
\end{figure}

Our system allows us to explore the spin-boson (SB) model in an ohmic bath. According to the SB model, the high frequency modes of the transmission line renormalize the bare qubit splitting $\Delta_0$ to \cite{spin-boson_rmp, juanjo_polaron}
\begin{equation}
\Delta=\Delta_0(p\Delta_0/\omega_C)^{\alpha_{\rm{SB}}/(1-\alpha_{\rm{SB}})}.
\end{equation}
$\alpha_{\rm{SB}}$ is the SB normalized coupling strength that is related to the spectral density of the environment $J(\omega)$. For an ohmic system such as our transmission line, $\alpha_{\rm{SB}}=J(\omega)/\pi\omega$. $\omega_C\gg\Delta_0$ is the cutoff frequency of the environment and $p$ is a constant of order 1. Up to $\alpha_{\rm{SB}}\sim0.5$, we identify $\alpha_{\rm{SB}}=\Gamma_1/\pi\Delta$. Above $\alpha_{\rm{SB}}\simeq0.5$ (or $\Gamma_1/\Delta\simeq\pi/2$) this relation becomes a lower bound for $\alpha_{\rm{SB}}$ (Supplementary Information). In Fig.~4(b) we plot the experimental qubit splittings $\Delta$ (circles). Using qubit junction dimensions extracted from SEM images of the device, we diagonalize the qubit Hamiltonian at each flux $\Phi_{\beta}$ (triangles) to give the bare qubit gaps $\Delta_0$. We then renormalize the calculated $\Delta_0$ using equation~3 and a value of $p=\exp(1+\gamma)\simeq4.8$, which is derived using an exponential cutoff model \cite{spin-boson_rmp, juanjo_polaron}.  $\gamma$ is the Euler constant. We find the best fit to the measured $\Delta$ using a cutoff of $\omega_C/2\pi=50$~GHz, which is consistent with characteristic system frequencies such as the plasma frequency of the qubit junctions and the superconducting gap. The agreement between the observed qubit splittings $\Delta$ and our estimates of the renormalized gaps is clear \cite{spin-boson_rmp, borja_usc, juanjo_polaron}. 

As a prelude to future work, we can place our results in the context of the SB model. The SB model defines three dynamical regimes for the qubit: underdamped ($\alpha_{\rm{SB}}<0.5$), overdamped ($1>\alpha_{\rm{SB}}>0.5$) and localized ($\alpha_{\rm{SB}}>1$). The connection between $\Gamma_1/\Delta$ and $\alpha_{\rm{SB}}$ allows us to draw the boundaries between these regimes in Fig.~4(a). We see that our tunable device enters well into the overdamped regime, and very possibly into the localized regime for $\beta<2$.  More detailed measurements of the dynamics of the device in these regimes could further confirm the predictions of the SB model. Suggestively, the strong reduction of the qubit response seen in Fig.~3(c) (leftmost data points in Fig.~4(a)) with a flat response as a function of frequency is consistent with simulations of classical double-well dynamics in the overdamped regime (in preparation, P. F.-D.).
 
We have presented measurements of superconducting flux qubits in 1D open transmission lines in regimes of interaction starting at strong coupling and ranging deeply into the ultrastrong coupling regime. In particular, we observed qubits with emission rates exceeding their own frequency, a clear indication of nonperturbative ultrastrong coupling. These results are very relevant for the study of open systems in the USC regime, opening the door to the development of a new generation of quantum electronics with ultrahigh bandwidth for quantum and nonlinear optics applications. The tunability of our system also makes it well-suited to the simulation of other quantum systems. In particular, we showed that the device can span the various transition regions of the SB model. With further development of our quantum circuit, the structure of the photon dressing cloud could also be directly detected, allowing the study of the physics of the Kondo model \cite{kondo} in a well-controlled setting. The ultrastrong coupling regime has other interesting intrinsic properties on its own, such as the entangled nature of the ground state.

Note added in proof: After acceptance of our paper, a related manuscript was published \cite{semba_dsc} showing similar results to this work using a resonator instead of a transmission line.

\section*{methods}

\noindent{\bf Device details and fabrication.}
We made the device with tuneable coupling by replacing the $\beta$-junction with a SQUID threaded by a flux $\Phi_{\beta}$. The tuneable coupling device then consists of two loops, the main loop that changes primarily the qubit magnetic energy through the flux $\Phi_{\epsilon}$ and the $\beta$-loop that changes the effective coupling to the transmission line through $\Phi_{\beta}$. Changing $\beta$ also modifies the minimum qubit splitting $\Delta$. In order to minimize this effect, we make the SQUID junctions asymmetric, which lowers the sensitivity of $\Delta$ to $\Phi_{\beta}$. Similar tuneable coupling architectures were already suggested in ref.~31. In the experiment, we sweep the global magnetic field, therefore simultaneously changing $\Phi_{\epsilon}$ and $\Phi_{\beta}$. The qubit spectrum shows minima near $\Phi_{\epsilon} \approx \Phi_0(1/2+n)$ with $\Phi_0 = h/2e$ the quantum of flux, $n$ being an integer (Supplementary Information). Here, different $n$ will correspond to different $\Phi_{\beta}$, leading to different coupling strengths. The loop areas $A_{\epsilon}/A_{\beta}$ are designed to have a large, incommensurate ratio, allowing the exploration of many different values of $\beta$. 

The fabrication methods used are based on those of ref.~27. The fabrication of devices starts by patterning the transmission line using optical lithography followed by an evaporation of 200~nm of aluminum. A gap in the transmission line is left to place the qubit in a second lithography stage. We pattern the qubit using an electron beam writer. Prior to the second aluminum evaporation an Ar milling step is applied to remove the native oxide on the first aluminum layer, guaranteeing optimal conduction between the two aluminum layers. The qubit is evaporated using double-angle shadow mask evaporation resulting in a total thickness of 105~nm. After the first shadow evaporation step, we oxidize the film with dynamical flow at $\sim0.01$~mbar for 7 minutes, yielding critical current densities of $\sim12~\mu$A/$\mu$m$^2$. The chip is then diced and the transmission line is wire-bonded to a printed circuit board connecting to the rest of the circuitry in our cryostat. 

The transmission line consists of a $6.5~$mm long on-chip coplanar waveguide with a center line and gaps 8~$\mu$m and $4~\mu$m wide, respectively, resulting in a 50~$\Omega$ characteristic impedance. Numerical simulations are run to verify the impedance of the circuit. We use a squared webbed ground to reduce superconducting vortex motion on the ground plane. 

\noindent{\bf Bounds on qubit emission rate.} 
The dependence of $r_0$ and $\Gamma_2$ on $n_{\mathrm{th}}$ shown below equation~(1) does not allow the independent extraction of all parameters, $\Gamma_1, \Gamma_{\varphi}, n_{\mathrm{th}}$ at each value of $\beta$. However, we can set bounds on $n_{\rm{th}}$. The lower bound case assumes no thermal excitations, therefore $n_{\rm th}=0$. If we instead set $\Gamma_{\varphi} = \Gamma_2(1-r_0(1+2n_{\mathrm{th}})^2)\ge0$, we identify an upper bound on the photon occupation number $n_{\rm{max}}\equiv(1/2)(1/\sqrt{r_0}-1)$. In Fig.~3(f), the values of $n_{\rm{max}}$ were extracted assuming $\Gamma_{\varphi}=0$. If we were to assume $\Gamma_{\varphi}/2\pi=17$~MHz as the nonthermal dephasing rate, extracted from the narrower linewidth of the device in Fig.~2(a) assuming $n_{\rm{th}}=0$, the resulting $n_{\rm{th}}$ would not differ significantly from $n_{\rm max}$. Now, bounds on $\Gamma_1=2\Gamma_2r_0(1+2n_{\mathrm{th}})$ can be set as $\Gamma_1(n_{\mathrm{th}}=0)$ and $\Gamma_1(n_{\mathrm{th}}=n_{\rm{max}})$ giving $2\Gamma_2r_0<\Gamma_1<2\Gamma_2\sqrt{r_0}$. The lower bound, $n_{\rm{th}}=0$, is close to the calculated value of $n_{\rm{th}}$ at the cryostat temperature of 10~mK for all qubit frequencies.

\noindent{\bf Spectroscopic analysis.} 
In all data shown, we use equation~(1) to simultaneously fit the real and imaginary parts of the transmission. Section S3 of the supplementary shows the full set of fitted resonances used in figures 3 and 4 of the main text. Note that the baseline is fixed to a normalized value of 1 and is not adjusted. The baseline value is itself determined by measuring the transmitted background when the qubit is flux-tuned away from the frequency band of interest.

\section*{Acknowledgements}
 We acknowledge financial support from NSERC of Canada, the Canadian Foundation for Innovation, the Ontario Ministry of Research and Innovation, Industry Canada, Canadian Microelectronics Corporation, EU FP7 FET-Open project PROMISCE, Spanish Mineco Project FIS2012-33022 and CAM Network QUITEMAD+. The University of Waterloo's Quantum NanoFab was used for this work.  
We thank A.~J.~Leggett and Anupam Garg for fruitful discussions, and M.~Otto, S.~Chang, A.~M.~Vadiraj and C.~Deng for help with device fabrication and with the measurement setups.

\pagebreak

\widetext

\setcounter{figure}{0}
\setcounter{equation}{0}
\makeatletter
\renewcommand{\thefigure}{S\@arabic\c@figure}
\renewcommand{\theequation}{S\arabic{equation}}

\makeatother

\section*{Supplementary material}
\section*{S0: Measurement setup}
\begin{figure}[!hbt]
\includegraphics{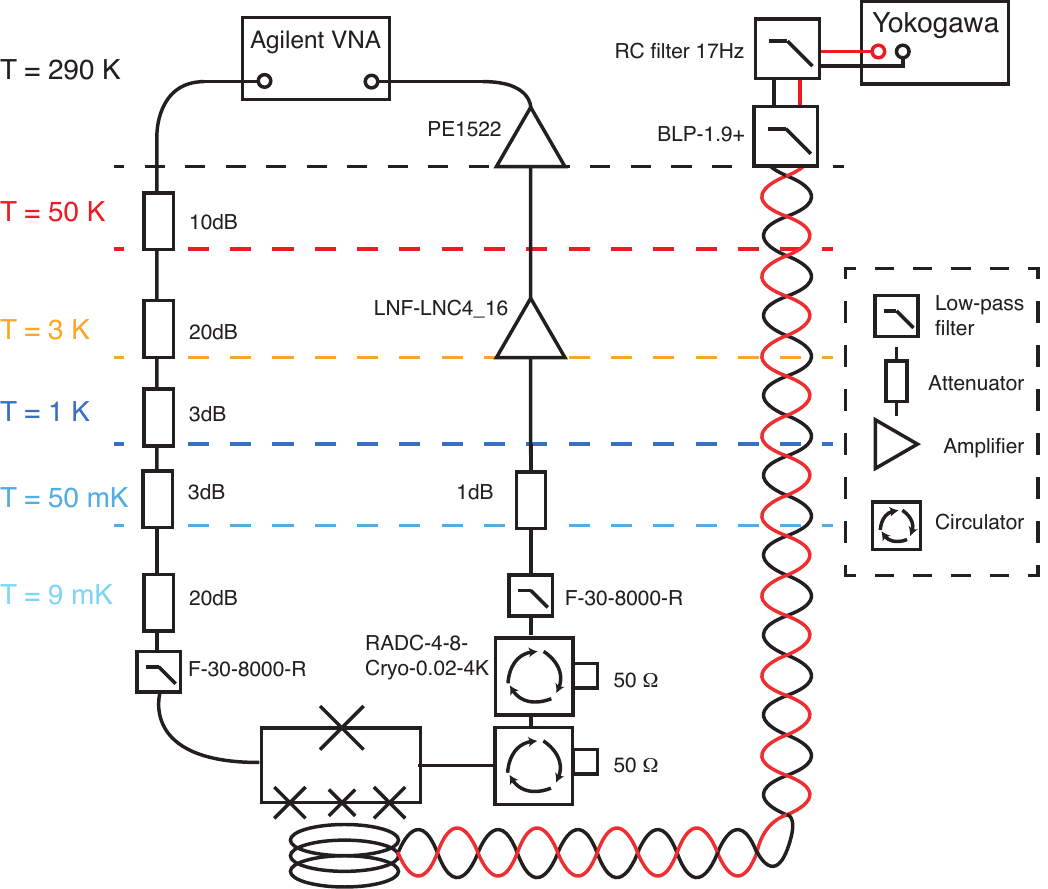}
\caption{\label{figs10}Schematic of the full circuit for transmission measurements.}
\end{figure}

Experiments on the device in Fig.~2(a) of the main text and the tunable device were performed in a dilution refrigerator with a base temperature of 9 mK where the chip is thermally anchored to. Our wiring is configured to measure both in reflection as well as in transmission using different input ports, although in this work we only show the transmission measurements. The optimal measurement bandwidth of the system is 4-8 GHz. The on-chip transmission line is followed by two circulators behind a cryogenic amplifier (see Fig.~\ref{figs10} for the full circuit diagram) anchored at 3.2~K with noise temperature of $\sim5$~K. We further amplify the signals at room temperature and digitize them using either a vector network analyzer or a spectrum analyzer. The device in Fig.~2(b) of the main text was characterized in a different dilution refrigerator with a base temperature of 15~mK, and having a similar wiring configuration as the one shown in Fig.~\ref{figs10} except for a larger nominal measurement bandwidth of 3-11 GHz.

\section*{S1: Qubit Hamiltonian and tunable coupling operator}
\begin{figure}[!hbt]
\centering
\includegraphics{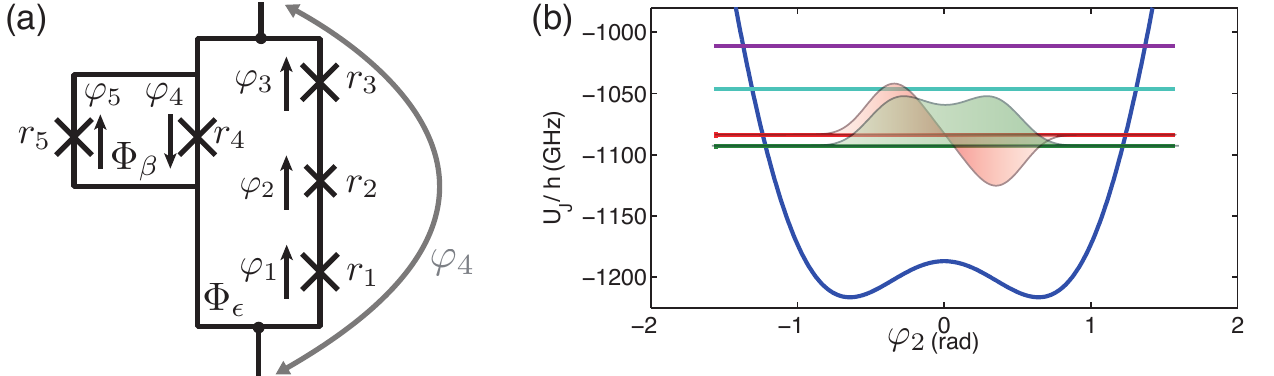}
\caption{\label{figs1}(a) Schematic of the circuit layout of a flux qubit sharing a SQUID-junction with a transmission line. The coefficients $r_1, r_2, r_3, r_4, r_5$ represent the size of the junctions. For the device used in the experiment, $r_2=0.62, r_4=2.6, r_1=r_3=r_5=1$. (b) Qubit potential with first four energy levels for $\Phi_{\beta}=0$, $\Phi_{\epsilon} = \Phi_0/2$ together with the wave functions of the lowest two levels for the ground (symmetric) and first excited state (antisymmetric). The parameters used are similar to the device with tunable coupling of the main text, $E_J/E_C\simeq70, E_J/h\simeq350~$GHz. Notice that the levels lie above the barrier as is usual for flux qubits with level splittings in the GHz range \cite{arkady_transition}.}
\end{figure}

The circuit layout of a flux qubit galvanically tunably coupled to a transmission line with a SQUID-loop shared between the two can be seen in Fig.~\ref{figs1}. 

The Lagrangian of the qubit can be written down by considering the fluxoid quantization condition on the separate loops:
\begin{align}
\varphi_1+\varphi_2+\varphi_3+\varphi_4+2\pi f_{\epsilon} &=0,\\
\label{fq2}\varphi_4+\varphi_5+2\pi f_{\beta} &=0,
\end{align}
where $f_{\epsilon} = \Phi_{\epsilon}/\Phi_0$, $f_{\beta} =\Phi_{\beta}/\Phi_0$ are the magnetic frustration in each loop. Using $\varphi_1,\varphi_2,\varphi_4$ as the independent degrees of freedom, the Lagrangian of the qubit reads \cite{devoret_houches}:
\begin{multline}\label{lagrangian}
\mathcal{L}(\varphi_1,\varphi_2,\varphi_4,\dot{\varphi_1},\dot{\varphi_2},\dot{\varphi_4})=\frac{\varphi_0^2C}{2}\left(r_1\dot{\varphi_1}^2+r_2\dot{\varphi_2}^2+(r_4+r_5)\dot{\varphi_4}^2+r_3(\dot{\varphi_1}+\dot{\varphi_2}+\dot{\varphi_4})^2\right)+\\E_J\big(r_1\cos\varphi_1+r_2\cos\varphi_2+r_4\cos\varphi_4+r_3\cos(-2\pi f_{\epsilon}-\varphi_1-\varphi_2-\varphi_4)+r_5\cos(-2\pi f_{\beta}-\varphi_4)\big).
\end{multline}
Here we defined the reduced flux quantum $\varphi_0=\Phi_0/2\pi$, $C$ is the capacitance of junction with size $r=1$, $E_J=I_C\varphi_0$. The canonical momenta $q_i=\partial\mathcal{L}/\partial\dot{\varphi}_i$ are related to the derivative of the conjugate phase operator:
\be
\begin{pmatrix}
\dot{\varphi}_1\\
\dot{\varphi}_2\\
\dot{\varphi}_4\end{pmatrix}
=\frac{1}{C\varphi_0^2}\frac{1}{\det(K)}
\begin{pmatrix}
r_3(r_4+r_5)+r_2(r_3+r_4+r_5) & -r_3(r_4+r_5) & -r_3r_2\\
-r_3(r_4+r_5) & r_3(r_4+r_5)+r_1(r_3+r_4+r_5) & -r_3r_1\\
-r_3r_2 & -r_3r_1 & r_2r_3+r_1(r_2+r_3)\end{pmatrix}
\begin{pmatrix}
q_1\\
q_2\\
q_4\end{pmatrix},
\ee
where $\det(K)=r_2r_3(r_4+r_5)+r_1(r_3(r_4+r_5)+r_2(r_3+r_4+r_5))$ is the dimensionless determinant of the capacitance matrix. We can now write down the Hamiltonian following a Legendre transformation $\mathcal{H}=\sum_iq_i\dot{\varphi}_i-\mathcal{L}$:
\begin{multline}\label{ham}
\mathcal{H}=
\frac{4E_C}{r_2r_3(r_4+r_5)+r_1(r_3(r_4+r_5)+r_2(r_3+r_4+r_5))}\big(n_1^2(r_3(r_4+r_5)+r_2(r_3+r_4+r_5))+\\
n_2^2(r_3(r_4+r_5)+r_1(r_3+r_4+r_5))+n_4^2(r_2r_3+r_1(r_2+r_3))-2n_1n_2r_3(r_4+r_5)-2n_4r_3(r_1n_2+r_2n_1)
\big)-\\E_J\big(r_1\cos\varphi_1+r_2\cos\varphi_2+r_4\cos\varphi_4+r_3\cos(-2\pi f_{\epsilon}-\varphi_1-\varphi_2-\varphi_4)+r_5\cos(-2\pi f_{\beta}-\varphi_4)\big).
\end{multline}
Here we defined the quantized charge operator $n_i=\hbar q_i$ as well as the charging energy $E_C=e^2/2C$. If we set $f_{\beta}=0$ the last term in the Josephson energy becomes $r_5\cos\varphi_4$, which combined with the $r_4$ term becomes an effective junction of size $(r_4+r_5)$. For $f_{\beta}=0.5$, the last term becomes $-r_5\cos\varphi_4$, which now leads to an effective junction of size $(r_4-r_5)$. Therefore we can tune the effective size of the coupling junction without affecting much of the rest of the qubit Hamiltonian. The different junction size will unavoidably lead to modifications of the qubit splitting.  

In order to diagonalize the Hamiltonian it is convenient to find its representation in the charge basis $\lbrace|n\rangle\rbrace$ where the Josephson terms have a simple expression, since \cite{devoret_houches} 
\be
\cos\varphi|n\rangle = \frac{e^{i\varphi}+e^{-i\varphi}}{2}|n\rangle=\frac{|n-1\rangle+|n+1\rangle}{2}.
\ee
The Josephson terms are therefore not represented by a closed Hilbert subspace in the charge basis. Therefore we need to restrict the number of charges between $-n_{\rm{max}}$ and $n_{\rm{max}}$ for each degree of freedom. Usually for $n_{\rm{max}} =10$ the error in the eigenenergies is less than 1\%. Fig~\ref{figs1}(b) shows the calculated qubit energies and wavefunctions for $f_{\beta}=0$ and $f_{\epsilon}=0.5$ using $n_{\rm{max}}=10$.

\subsection*{S1.1: Energy levels and Crosstalk}
For a given set of fluxes $(f_{\epsilon},f_{\beta})$ we can find the eigenenergies and eigenstates of the qubit. For all calculations shown in this section we take the values close to the experiment with the tunable coupling device $r_1=r_3=1.0, r_2=0.6, r_4=1.0, r_5=2.6$, $E_J/E_C=70$, $E_J/h=300$~GHz, $n_{\rm max}=10$. The areas of the two qubit loops are seen to be $A_{\epsilon}/A_{\beta}\simeq8.3$, which agree with the data as seen in the calculations of Fig.~4(b) of the main text. In order to reproduce the experimental spectra we sweep the flux in the $\epsilon$-loop and assume the flux in the beta loop to be proportional to it, $\Phi_{\beta}=\Phi_{\epsilon}/8.3$.
\begin{figure}[!hbt]
\centering
\includegraphics{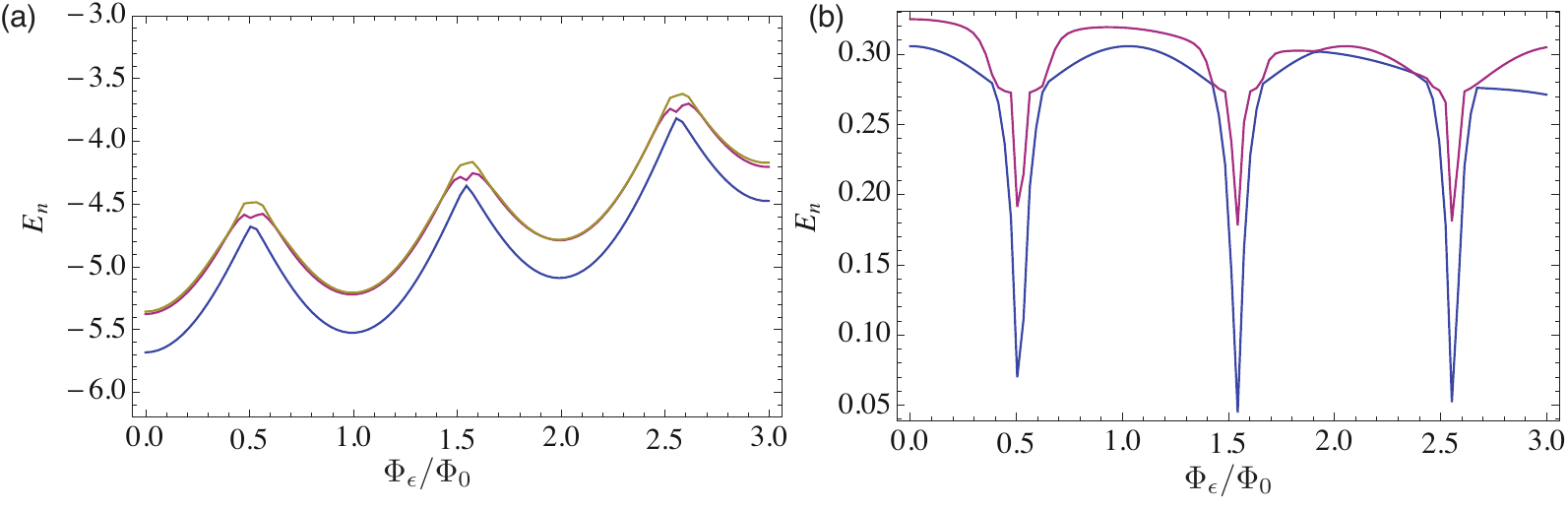}
\caption{\label{figs2}(a) Calculated three lowest energy levels of the Hamiltonian in Eq~(\ref{ham}) as function of $\Phi_{\epsilon}$ taking into account that $\Phi_{\beta}=\Phi_{\epsilon}/8.3$. The energies are in units of $E_J$. (b) Energy differences with respect to the ground state energy. Notice that qubit symmetry points are not falling on top of $\Phi/\Phi_0=0.5$ due to the interference between the two qubit loops.}
\end{figure} 

\begin{figure}[!hbt]
\centering
\includegraphics{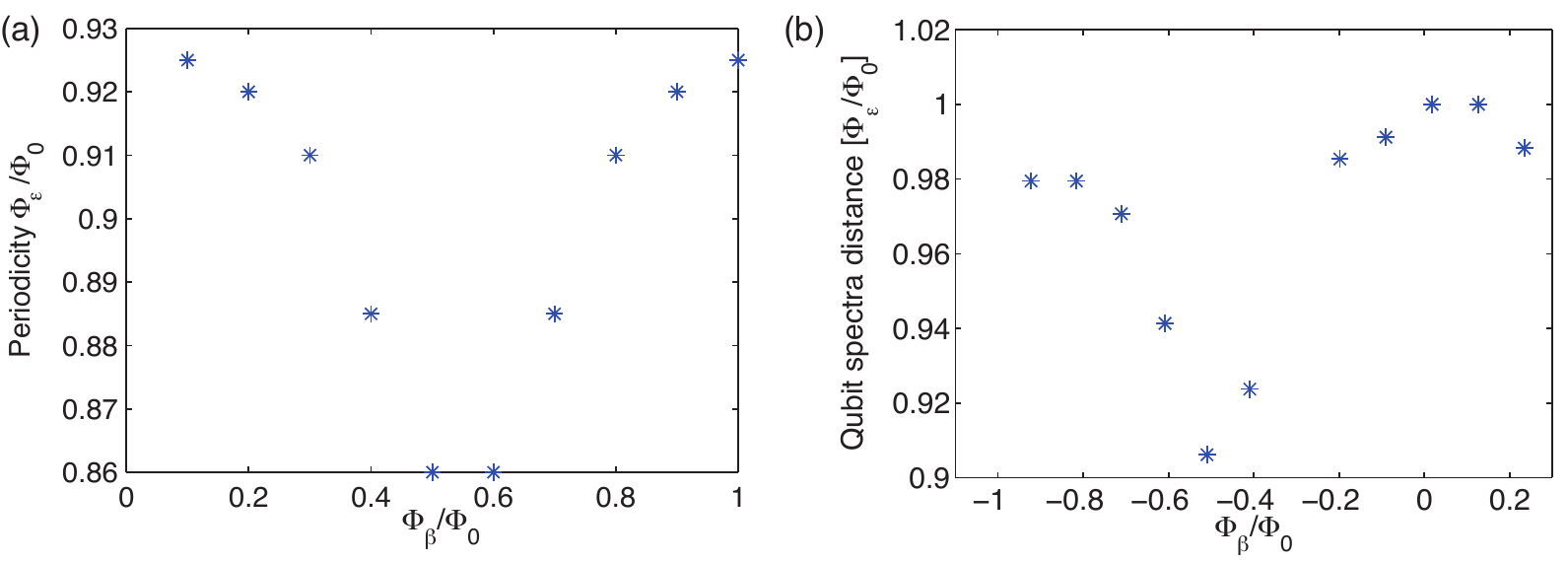}
\caption{\label{figs3}(a) Calculated difference in consecutive qubit symmetry points as function of flux in $\epsilon$-loop, using Eqs.~(\ref{ftilde1}),~(\ref{ftilde2}). (b) Experimentally measured distance between qubit periods. The relative change of period $\sim10\%$ for both plots (a) and (b) agrees quite well. The difference in (b) at $\Phi_{\beta}/\Phi_0=0$ and $\Phi_{\beta}/\Phi_0\simeq-1$ could be attributed to small flux drifts, given that the sweep is over many periods of flux for the qubit $\epsilon$-loop.}
\end{figure}

The resulting spectra in Fig.~\ref{figs2} clearly show a lack of periodicity, as would be expected for a qubit with no SQUID-loop. The qubit symmetry points do not agree with $\Phi_{\epsilon} = \Phi_0(1/2+n)$, $n$ being an integer. The difference is due to the interference between the two qubit loops. 
The potential energy terms related to the applied fluxes are
\be
r_4\cos(-2\pi f_{\epsilon}-\varphi_1-\varphi_2-\varphi_3)+r_5\cos(-2\pi (f_{\beta}+f_{\epsilon})-\varphi_1-\varphi_2-\varphi_3).
\ee
In analogy with the usual flux qubit potential \cite{hans_sci}, we can rewrite these terms as an effective new Josephson term $\tilde{U}$ with effective critical current $\tilde{I}$ and effective flux $\tilde{f}$ as $\tilde{U}/E_J=-\tilde{I}\cos(-\varphi_1-\varphi_2-\varphi_4 + 2\pi\tilde{f})$. Expanding the cosine terms, we can relate $\tilde{f}$ and $\tilde{I}$ with the rest of parameters: 
\begin{align}
\label{tildepot}&-\tilde{U}/E_J = \tilde{I}(\cos\varphi_{\Sigma}\cos2\pi\tilde{f} - \sin\varphi_{\Sigma}\sin2\pi\tilde{f}),\\
\label{tildepot2}&= \cos\varphi_{\Sigma}(r_4\cos2\pi f_{\epsilon}+r_5\cos(2\pi(f_{\epsilon}+f_{\beta}))-\sin\varphi_{\Sigma}(r_4\sin2\pi f_{\epsilon} + r_5\sin2\pi(f_{\epsilon}+f_{\beta})),
\end{align}
where $\varphi_{\Sigma}=\varphi_1+\varphi_2+\varphi_3$. In particular, at the symmetry point $\tilde{f} = 1/2$ which cancels the second term in Eq.~(\ref{tildepot}). This implies that the term multiplying $\sin\varphi_{\Sigma}$ has to be zero, leading to a transcendental equation to obtain the location of all symmetry points:
\be\label{ftilde1}
-\frac{r_4}{r_5}=\frac{\sin2\pi f_{\epsilon}}{\sin2\pi(f_{\epsilon}+f_{\beta})}.
\ee
In addition to equation (10) we also impose the condition $\cos(2 \pi \tilde{f}) = -1$, that is
\be\label{ftilde2}
-1 = r_4\cos2\pi f_{\epsilon} + r_5\cos2\pi(f_{\epsilon}+f_{\beta}).
\ee
We calculate the difference between consecutive qubit symmetry points in Fig.~\ref{figs3}(a) over a period of $\Phi_{\beta}$. The cosine-like modulation clearly shows the interference between the two loops. The experimentally measured difference in periodicity of the qubit symmetry points is plotted in Fig.~\ref{figs3}(b). A modulation of the periodicity is also clear. The relative change of periodicity of $\sim10\%$ agrees with the prediction of Eqs.~(\ref{ftilde1}),~(\ref{ftilde2}). Fig.~\ref{figs3}(b) is scaled to the value at $\Phi_{\beta}/\Phi_0=0$. Fig.~\ref{figs3}(a) is scaled such that ``1" in the vertical axis would correspond to $\Phi_{\epsilon}/\Phi_0=0.5$. 

The qubit Lagrangian shown here does not include geometric capacitance between islands and to ground. We have inspected the effect of those terms and found less than 10\% variation in the qubit frequency. 

\subsection*{S1.2: Coupling operator}
\begin{figure}[!hbt]
\centering
\includegraphics{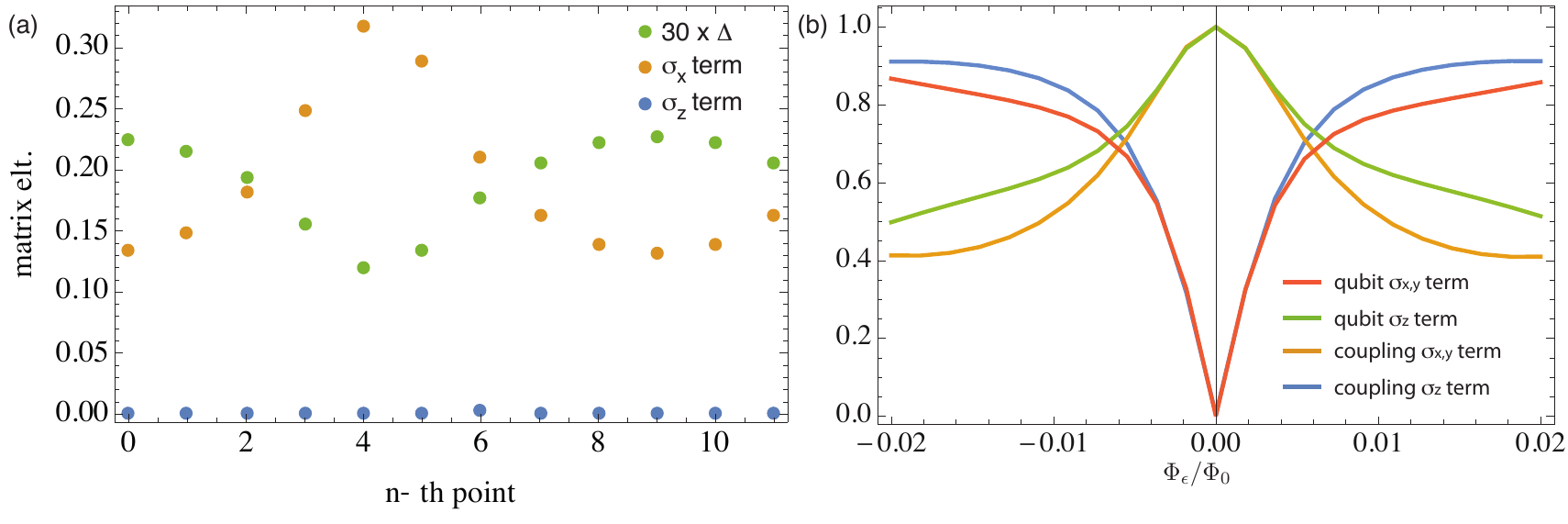}
\caption{\label{figs4}(a) Calculated phase operator in the qubit basis at the symmetry point for different periods of $\Phi_{\epsilon}$. The operator has only $\sigma_x$ component (orange) and no $\sigma_z$ component (blue). The qubit splitting $\Delta$ is also plotted, its value changing approximately a factor of 2 for the parameters of the device. (b) Calculated qubit Hamiltonian terms and coupling operator terms as function of $\Phi_{\epsilon}$ for fixed $\Phi_{\beta}=0$. Away from the symmetry point the qubit Hamiltonian rotates from $\sigma_z$ and starts to acquire a $\sigma_{x,y}$ component. The coupling operator follows closely the qubit Hamiltonian rotation. Therefore near the symmetry point the coupling operator rotates due to the qubit basis rotation as function of magnetic flux $\Phi_{\epsilon}$ and not due to other terms in the Hamiltonian.}
\end{figure}

As shown in Ref.~\cite{borja_usc}, the coupling strength of a flux qubit sharing a junction with a resonator or a transmission line is given by the modulus of the matrix element of the phase operator across the coupling junction $|\langle1|\hat{\varphi}_i|0\rangle|$. In the circuit of Fig.~\ref{figs1} this corresponds to the phase across $\hat{\varphi}_4$. The coupling operator can be expressed in the qubit basis states using that the representation in the charge basis of the phase operator is
\be
\langle n|\hat{\varphi}|m\rangle = \frac{1}{2\pi}\int_{-\pi}^{\pi}\hat{\varphi} e^{-i(m-n)\hat{\varphi}}\mathrm{d}\hat{\varphi} = 
\begin{cases} 0 & \text{if } m = n,\\
-i\frac{(-1)^{(m-n)}}{m-n} & \text{if } m \ne  n.
\end{cases}
\ee
The limits of integration fall within a unit cell of the qubit potential. The qubit eigenstates can be represented in the basis of charge states $\displaystyle|g\rangle = \sum_{n_1,n_2,n_4=-n_{max}}^{n_{max}}c_{n_1,n_2,n_4}|n_1,n_2,n_4\rangle$. Therefore the matrix elements of the phase operator in the qubit basis $\lbrace|g\rangle,|e\rangle\rbrace$ for arbitrary states $|M\rangle,|N\rangle$ look like:
\be\label{ph_op}
\langle M|\hat{\varphi}_4|N\rangle = \sum_{n_1,n_2,n_4}^{n_{\rm max}}\sum_{n_1',n_2',n_4'}^{n_{\rm max}}c_{n_1,n_2,n_4}^{\ast}c_{n_1',n_2',n_4'}\left[-i\frac{(-1)^{(n_4'-n_4)}}{n_4'-n_4}\delta_{n_1,n_1'}\delta_{n_2,n_2'}\right].
\ee
We use the representation of the phase operator in the qubit basis to obtain the different components of the qubit operator. We want only transverse coupling $\sigma_x$ with matrix element $\langle0|\hat{\varphi}_4|1\rangle\equiv\varphi_4$, and not $\sigma_z$ terms that would otherwise induce dephasing in the qubit from the line. Using equation (\ref{ph_op}) we can compute the form of the operator, shown in Fig.~\ref{figs4}

Clearly, the coupling operator only has $\sigma_x$ component at the symmetry point while its magnitude increases by approximately a factor of 2.6, as expected due to the modulation of the size of the $\beta$-junction. The qubit gap is also modulated as expected due to the effective change in size of $\beta$. The change is approximately of a factor of 2. Therefore the normalized coupling $\Gamma/\Delta\sim|\varphi_{\beta}|^2$ (see section~S7) increases by  a factor of $\sim7$ from $\Phi_{\beta}=0$ to $\Phi_{\beta}=\Phi_0/2$. On Fig.~\ref{figs4}(b) we also calculate the terms of the qubit Hamiltonian and coupling operator near $\Phi_{\epsilon}=\Phi_0$ for fixed $\Phi_{\beta}=0$. Clearly the coupling rotates from $\sigma_x$ to $\sigma_z$ as the qubit Hamiltonian rotates from $\sigma_z$ to $\sigma_x$. Therefore the rotation of the coupling operator is exclusively due to the qubit Hamiltonian rotation and not from other terms. Similar rotations of the coupling operator between qubits and transmission lines or resonators were already identified in \cite{borja_prl}.

\section*{S2: Scattering rates at finite temperature}
Following from \cite{borja_scatt}, the general definition of the master equation is
\be
\dot{\rho}(t) = -\frac{i}{\hbar}[\mathcal{H},\rho(t)] + \frac{1}{2}\sum_n[2C_n\rho(t)C_n^{\dag}-\rho(t)C_n^{\dag}C_n-C_n^{\dag}C_n\rho(t)],
\ee
where the $C_n=\sqrt{\gamma}A_n$ are the jump operators, with $\gamma$ the rate for each process with collapse operator $A_n$. Here $\mathcal{H}$ is the free Hamiltonian and contains the qubit part truncated to two levels, $\mathcal{H}_{\rm{qb}}/\hbar=\Delta\sigma_x/2-\epsilon\sigma_z/2$ and the external driving $\mathcal{H}_d/\hbar=\Omega\sin(\omega_dt)\sigma_z$. For a qubit, the decay operator is $\sigma_-$ while the excitation operator is $\sigma_+$. In terms of components, the diagonal decay terms of the master equation read
\be
\dot{\rho}_{ii} = \sum_{j\ne i}(\Gamma_{ji}\rho_{jj} - \Gamma_{ij}\rho_{ii}).
\ee
Here $\Gamma_{eg} = \Gamma_1(1+\nth)$ is the relaxation rate and $\Gamma_{ge} = \nth\Gamma_1$ the excitation rate. $n_{\rm th}$ is the expectation value of photon number for a thermal state in equilibrium with a bath at temperature $T$:
\[n_{\rm th}=\frac{1}{e^{\hbar\omega/k_BT}-1}.\]The off-diagonal decay terms take the form (for $i\ne j$)
\be
\dot{\rho}_{ij} = -\gamma_{ij}\rho_{ij},
\ee
where the decoherence rates are $\gamma_{eg} = \Gamma_{\varphi}+\frac{1}{2}(\Gamma_{eg}+\Gamma_{ge}) = \Gamma_{\varphi}+(\Gamma_1/2)(1+2\nth)\equiv\Gamma_2$, with $\Gamma_{\varphi}$ the pure dephasing rate.

Explicitly, the decay equations for the four components of the density matrix now look:
\begin{align}
\dot{\rho}_{ee}& = \Gamma_{ge}\rho_{gg} - \Gamma_{eg}\rho_{ee} = \nth\Gamma_1(\rho_{gg}-\rho_{ee})-\Gamma_1\rho_{ee},\\
\dot{\rho}_{gg}& = \Gamma_{eg}\rho_{ee} - \Gamma_{ge}\rho_{gg} = \nth\Gamma_1(\rho_{ee}-\rho_{gg})+\Gamma_1\rho_{ee},\\
\dot{\rho}_{eg}& = -\gamma_{eg}\rho_{eg} = -[\Gamma_{\varphi}+\Gamma_1(1+2\nth)/2]\rho_{eg}=-\Gamma_2\rho_{eg},\\
\dot{\rho}_{ge}& = -\gamma_{ge}\rho_{ge} = -[\Gamma_{\varphi}+\Gamma_1(1+2\nth)/2]\rho_{ge}=-\Gamma_2\rho_{ge}.
\end{align}

The free evolution terms given by the commutator $[\mathcal{H},\rho(t)]$ can be easily computed in the rotating frame of the drive frequency $\omega_d$, under the rotating-wave approximation, where
\be
\mathcal{H}/\hbar = -\delta\omega\sigma_z/2+\Omega\sigma_x/2.
\ee 
The detuning is defined as $\delta\omega = \omega_d- \omega_{\rm{qb}}$, $\omega_{\rm{qb}}=\sqrt{\Delta^2+\epsilon^2}$ is the qubit energy splitting in units of angular frequency. $\epsilon$ is the magnetic field energy controlled by $\Phi_{\epsilon}$ (Fig.~\ref{figs1}(a)). The full equation of motion for all components of the density matrix are:
\begin{align}
\label{me1}\dot{\rho}_{ee}& = -\frac{i\Omega}{2}(\rho_{ge}-\rho_{eg}) + \nth\Gamma_1(\rho_{gg}-\rho_{ee})-\Gamma_1\rho_{ee},\\
\label{me2}\dot{\rho}_{gg}& = +\frac{i\Omega}{2}(\rho_{ge}-\rho_{eg}) + \nth\Gamma_1(\rho_{ee}-\rho_{gg})+\Gamma_1\rho_{ee}=-\dot{\rho}_{ee},\\
\label{me3}\dot{\rho}_{eg}& = -\frac{i\Omega}{2}(\rho_{gg}-\rho_{ee})-i\delta\omega\rho_{eg}-\Gamma_2\rho_{eg},\\
\label{me4}\dot{\rho}_{ge}& = +\frac{i\Omega}{2}(\rho_{gg}-\rho_{ee})+i\delta\omega\rho_{ge}-\Gamma_2\rho_{ge}.
\end{align}

Now we want to find the steady-state populations of the qubit, $\dot{\rho}=0$. The off-resonant terms are related by
\be\label{eq13}
\rho_{eg}(i\delta\omega+\Gamma_2) = \rho_{ge}(i\delta\omega-\Gamma_2).
\ee
Adding Eqs.~(\ref{me1}),~(\ref{me4}),
\be\label{eq14}
\Gamma_1\rho_{ee} = -\rho_{ge}\left(\frac{i\Omega}{2}\right)\left(\frac{2\Gamma_2}{\Gamma_2+i\delta\omega}+(\Gamma_2-i\delta\omega)\Gamma_1\nth\left(\frac{2}{\Omega}\right)^2\right).
\ee
Using that Tr$(\rho)=1=\rho_{ee}+\rho_{gg}$, Eq.~\ref{me2} can be rewritten as
\be\label{eq15}
\rho_{ee} = \frac{1}{\Gamma_1(1+2\nth)}\left(\Gamma_1\nth - i\frac{\Omega}{2}(\rho_{ge}-\rho_{eg})\right).
\ee
Combining Eqs.~(\ref{eq13})-(\ref{eq15}) directly gives the solution for $\rho_{ge}$:
\be
\rho_{ge}=\frac{i\Omega}{2}\frac{\Gamma_1(\Gamma_2+i\delta\omega)}{\Gamma_2\Omega^2+\Gamma_1(\Gamma_2^2+\delta\omega^2)(1+2\nth)}.
\ee
Using Eq.~(\ref{eq13}) provides $\rho_{eg}$:
\be
\rho_{eg}=-\frac{\Omega}{2\Gamma_2}\frac{i+\delta\omega/\Gamma_2}{(1+(\delta\omega/\Gamma_2)^2)(1+2\nth)+\Omega^2/(\Gamma_1\Gamma_2)}.
\ee
Following \cite{astafiev_fluor}, the reflection coefficient is defined as $r\equiv -i(\Gamma_1/\Omega)\langle\sigma_-\rangle$. It is easy to see that $\langle\sigma_-\rangle=\rho_{eg}$. 

Therefore adding finite temperature to the system modifies the scattering parameters as follows:

\be\label{rsc} r =r_0\frac{(-1+i\delta\omega/\Gamma_2)}{1+\left(\frac{\delta\omega}{\Gamma_2}\right)^2+\frac{\Omega_R^2}{\Gamma_1\Gamma_2}},\ee
with $r_0 \equiv \Gamma_1/(2\Gamma_2(1+2n_{\mathrm{th}}))$ and $\Omega_R\equiv\Omega/\sqrt{1+2n_{\rm th}}$. The form of Eq.~(\ref{rsc}) is the same as the usual reflection coefficient if $\nth=0$. Therefore the fitted values for $r_0$ and $\Gamma_2$ are going to be independent of temperature, the difference will appear in $\Gamma_1$ and $\Gamma_{\varphi}$. The transmission coefficient will be given by $t = 1+r$
\be
t  = 1+r =
\frac{1+(\delta\omega/\Gamma_2)^2+r_0(-1+i\delta\omega/\Gamma_2)+\frac{\Omega_R^2}{\Gamma_1\Gamma_2}}{1+(\delta\omega/\Gamma_2)^2+\frac{\Omega_R^2}{\Gamma_1\Gamma_2}}\simeq\frac{1+(\delta\omega/\Gamma_2)^2+r_0(-1+i\delta\omega/\Gamma_2)}{1+(\delta\omega/\Gamma_2)^2},
\ee
where the last step assumed weak driving $\Omega_R^2\ll\Gamma_1\Gamma_2$. The resulting expression is the function used to fit the data, equation 2 in the main article. The minimum of transmission on-resonance in this case is 
\[t_{\rm min}=\frac{4\Gamma_1n_{\rm th}+2\Gamma_{\varphi}(1+2n_{\rm th})+4\Gamma_1n_{\rm th}^2}{(1+2n_{\rm th})(\Gamma_1+2\Gamma_{\varphi}+2\Gamma_1n_{\rm th})}.\]
Setting $n_{\rm th}=0$ one restores the result of $t_{\rm min}(n_{\rm th}=0)=\Gamma_{\varphi}/\Gamma_2=\Gamma_{\varphi}/(\Gamma_1/2+\Gamma_{\varphi})$.

\begin{figure}[!hbt]
\centering
\includegraphics{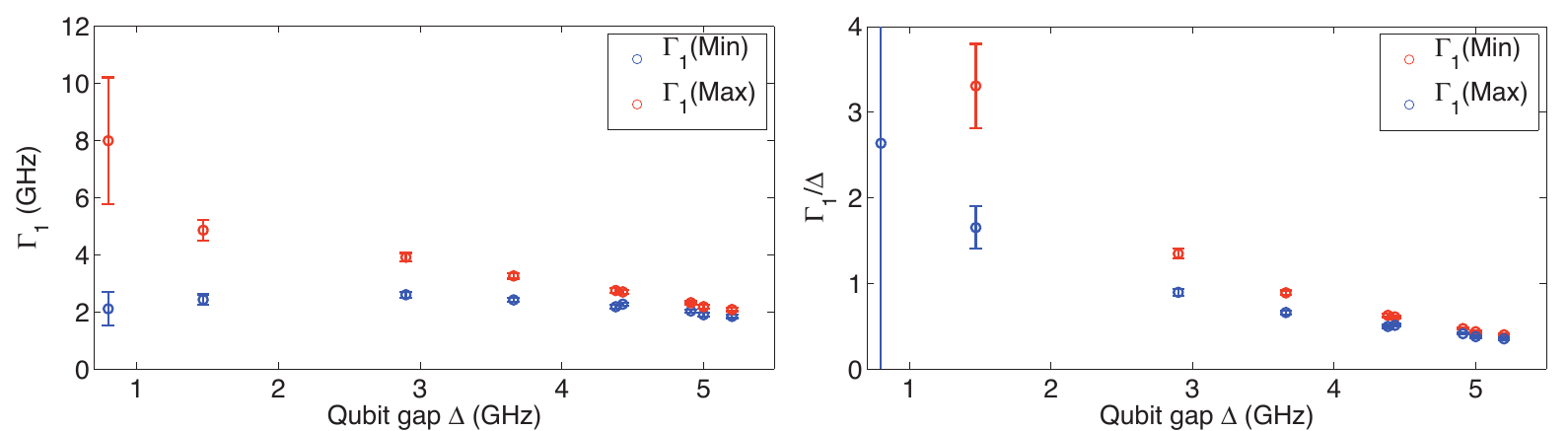}
\caption{\label{figs9}(a) $\Gamma_1$ as function of qubit gap $\Delta$, which corresponds to different coupling rates. (b) $\Gamma_1/\Delta$ as function of $\Delta$.}
\end{figure}

The extracted values of $\Gamma_1$ from the experiment can be then bound assuming no thermal photons (lower bound) or the maximum number of photons allowed if $\Gamma_{\varphi}=0$ (upper bound), as seen in Fig.~\ref{figs9}

\section*{S3: Combined fitting of Re($T$) and Im($T$)}
\begin{figure}[!hbt]
\centering
\includegraphics{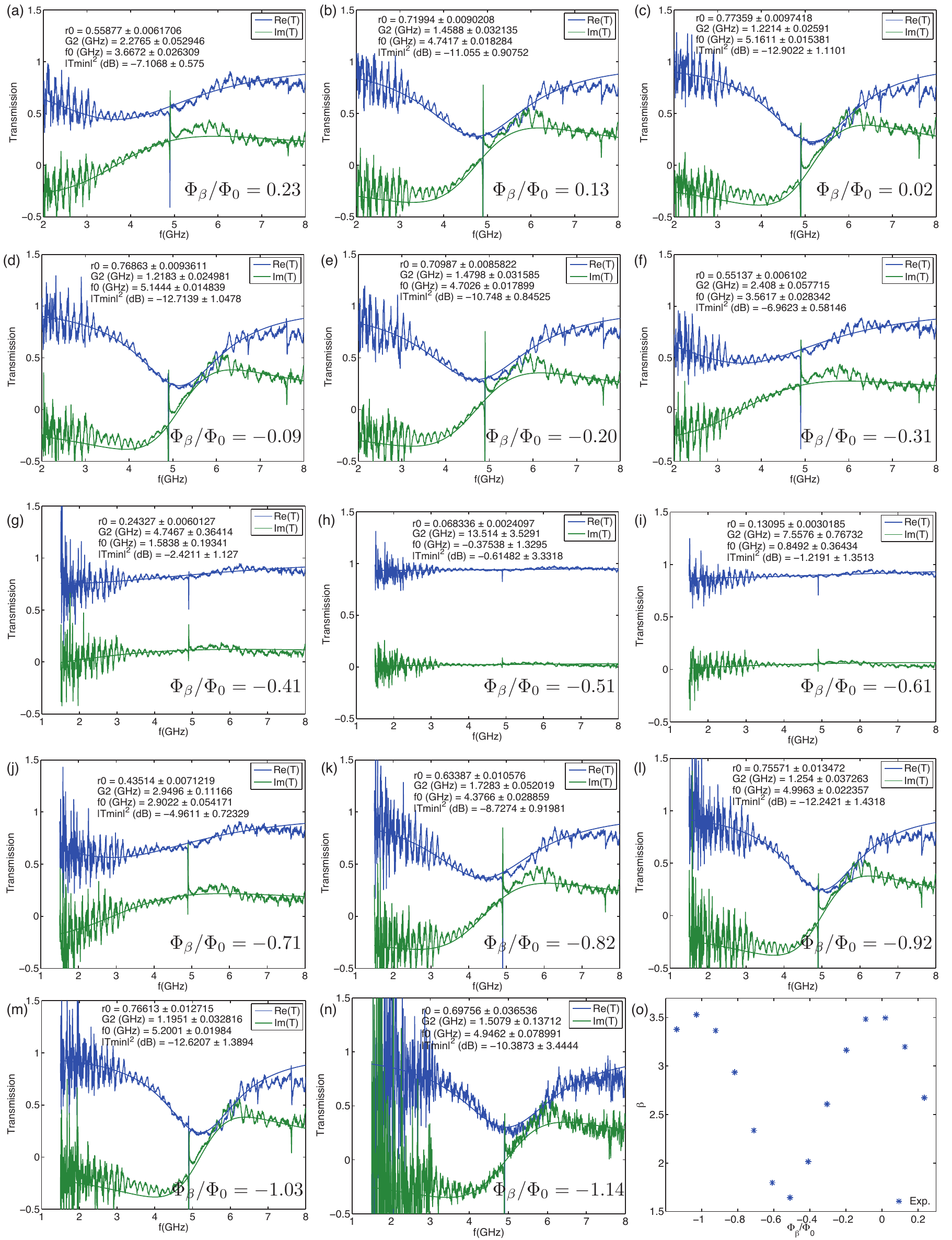}
\caption{\label{figs5}(a)-(n) Combined fits corresponding to data in Fig.~3 from the main text. (o) Effective size of junction $\beta(\Phi_{\beta})$.}
\end{figure}

\begin{figure}[!hbt]
\centering
\includegraphics{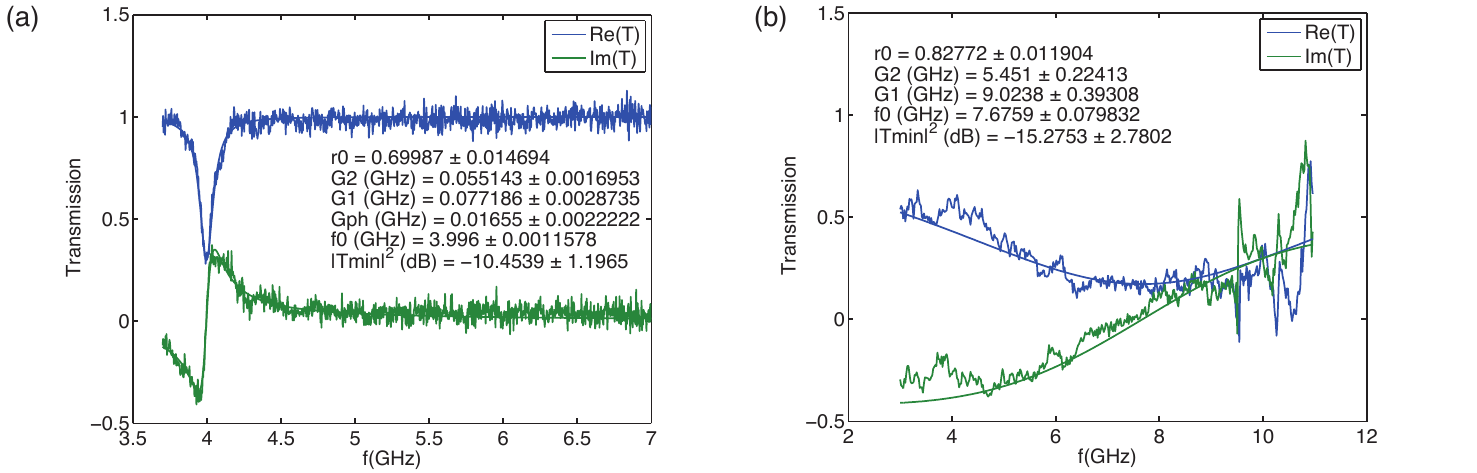}
\caption{\label{figs6}Combined fits from devices with fixed coupling from Fig.~2 of the main text (a) $\beta=3.5$, (b) $\beta$=1.8.}
\end{figure}

The fits in figures 2, 3 of the main text are performed simultaneously on both the real and imaginary parts. We show here the total fitted transmission components. Fig.~\ref{figs5} corresponds to the fits of the tunable device while Fig.~\ref{figs6} corresponds to the devices with fixed coupling. As explained in the main text, even though the extracted emission rates correspond to the regime where the rotating-wave approximation (RWA) is not valid, by rotating the basis of the system Hamiltonian using a polaron transformation \cite{juanjo_polaron} the functional form of the real and imaginary parts of the transmission follow the same analytical form as the RWA case, with a renormalized qubit splitting $\Delta$ and emission rate $\Gamma_1$ instead. Due to the fact that most data is taken below the optimal bandwidth of our amplifier and circulators below 4~GHz, the quality of the fits degrades as the system enters the regime $\Gamma_1>\Delta$ (plots (g), (h), (i) in Fig.~\ref{figs5}). The only relevant parameters extracted are $r_0$ and $\Gamma_2$. As explained in the main text, $r_0$ and $\Gamma_2$ are enough to set bounds on $\Gamma_1$ and the effective temperature of the system.

\section*{S4: Estimates of dephasing rate}
The SQUID loop in the qubit with tunable coupling may be an additional source of decoherence, especially dephasing noise since fluctuations in the flux $\Phi_{\beta}$ will directly convert into fluctuations of the qubit gap $\Delta$. We have assumed in the main text that the enhanced linewidth of the qubit is due to thermal effects. The justification is made here where we put bounds on possible sources of dephasing. 

The sensitivity of the qubit gap as function of $\Phi_{\beta}$ can be estimated by using the modulation of the qubit gap curve on Fig.~4 of the main text. An upper bound for the flux sensitivity is $\rm{d}\Delta/\rm{d}\Phi_{\beta} \sim 7.4~\rm{GHz}/\Phi_0$. Comparing this number to the sensitivity of the qubit to flux $\Phi_{\epsilon}$, $\rm{d}\omega_{\rm{qb}}/\rm{d}\Phi_{\epsilon}\sim5~\rm{GHz}/(2.5\times10^{-3}\Phi_0)\sim2\times10^{3}~\rm{GHz}/\Phi_0$ we can see that sensitivity to flux noise in the $\beta$-loop is negligible.

Another possible source of flux noise would be through the qubit renormalization frequency as predicted by the spin-boson model. The model predicts \cite{spin-boson_rmp} that the splitting of a two-level system in a bath of oscillators will be adiabatically renormalized to $\Delta = \Delta_0(\Delta_0/\omega_C)^{\alpha_{\rm{SB}}/(1-\alpha_{\rm{SB}})}$, where $\omega_C$ is the cutoff frequency of the environment and $\Delta_0$ is the bare qubit gap. Since $\alpha_{\rm{SB}} = \Gamma_1/\pi\Delta$ (see section S6) and therefore both $\Gamma_1(\Phi_{\beta}), \Delta(\Phi_{\beta})$ depend on $\Phi_{\beta}$, fluctuations in $\Phi_{\beta}$ may lead to fluctuations in $\Delta$. The sensitivity can be calculated:
\be
\frac{\partial\Delta}{\partial\Phi_{\beta}}=\left(\frac{\Delta_0}{\omega_C}\right)^{\frac{\alpha_{\rm{SB}}}{1-\alpha_{\rm{SB}}}}\left[ 2\frac{\mathrm{d}\Delta_0}{\mathrm{d}\Phi_{\beta}}+\ln(\Delta_0/\omega_C)\frac{1}{(1-\alpha_{\rm{SB}})^2}\frac{\Gamma_1'\Delta_0 - \Gamma_1\Delta_0'}{\Delta_0}\right]\simeq\left(\frac{\Delta_0}{\omega_C}\right)^{\frac{\alpha_{\rm{SB}}}{1-\alpha_{\rm{SB}}}}\frac{\mathrm{d}\Delta_0}{\mathrm{d}\Phi_{\beta}}\left[2+\ln(\Delta_0/\omega_C)\frac{1+\alpha_{\rm{SB}}}{(1-\alpha_{\rm{SB}})^2}\right].
\ee
Here we used that in our experiment (Fig.~\ref{figs9}(a)) $\Gamma_1'\equiv\rm{d}\Gamma_1/\rm{d}\Phi_{\beta}\approx-\Delta_0'\equiv-\rm{d}\Delta_0/\rm{d}\Phi_{\beta}$. The highest sensitivity occurs for $\alpha_{\rm{SB}}=1/2$ where $\partial\Delta/\partial\Phi_{\beta}\simeq-2(\rm{d}\Delta_0/\rm{d}\Phi_{\beta})$, assuming a worst case $\Delta/\omega_C\sim1/10$. In the main text we find $\Delta/\omega_C\sim1/15$ as the worst case. Therefore this source of dephasing is also negligible.

\section{S5: Temperature sweeps}
\begin{figure}[!hbt]
\centering
\includegraphics[width = 0.9\columnwidth]{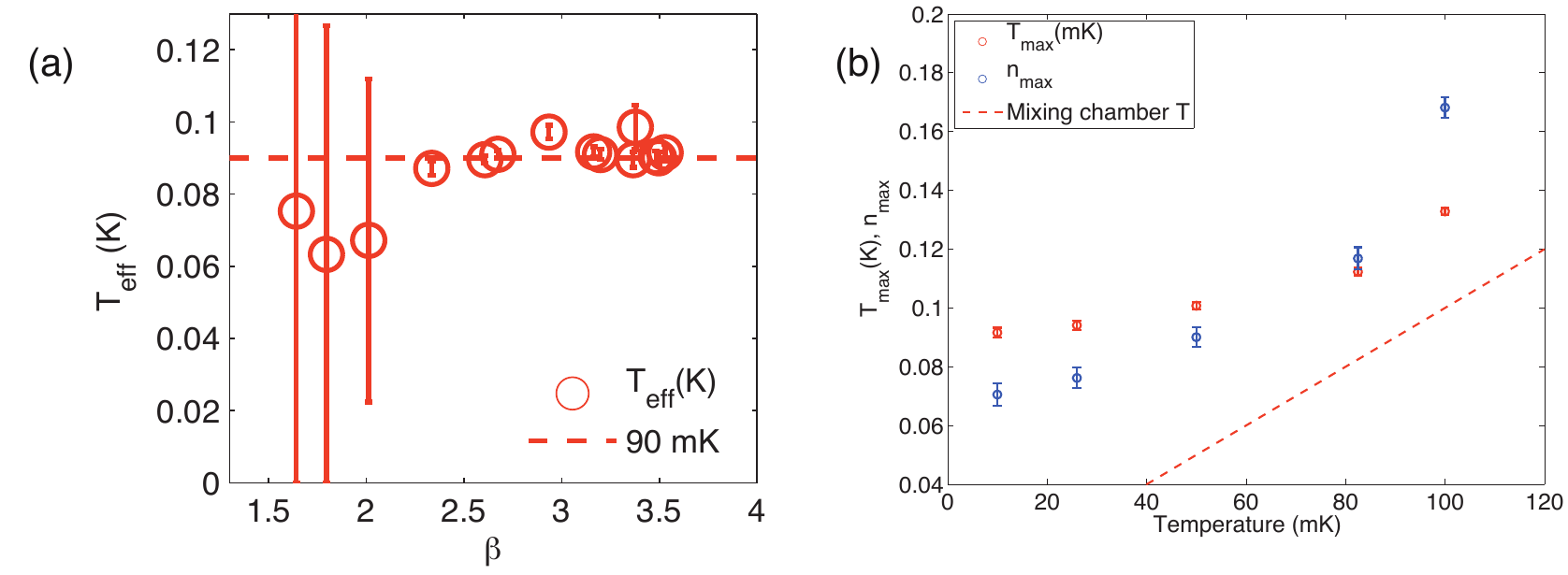}
\caption{\label{figs7}(a) Calculated effective temperature from all qubit spectra in Fig.~3 of the main text. (b) Effective thermal photon number $n_{\rm{max}}$ (blue dots) and effective temperature (red dots) extracted from spectroscopy fits of qubit with tunable coupling resonances at bias flux where qubit has highest frequency $\Delta/2\pi\sim5.2~$GHz (Fig.~3(a) main text). Above cryostat temperatures of $\sim80~$mK the effective qubit temperature increases at the same rate as the cryostat. Both sets of measurements support an maximum effective temperature $T_{\rm{eff}}$ seen by the qubit of 90~mK.}
\end{figure}

We want to establish more solid bounds on the maximum effective temperature $T_{\rm{eff}}=90~$mK extracted from the fits of qubit spectra at different flux values, shown in Fig.~\ref{figs7}(a), which complements the inferred $n_{\rm{max}}$ in Fig.~3(f) of the main text. Here, we study the resonance on Fig.~3(a) from the main text, where the qubit frequency is highest, as function of the base temperature of our cryostat, which is where our device is thermalized to. 

In Fig.~\ref{figs7}(b) we show the extracted maximum photon number $n_{\rm{max}} = (1/2)(r_0^{-1/2}-1)$ and the corresponding effective temperature $T_{\rm{eff}} = (\hbar\Delta/k_B)\ln(1+n_{\rm{max}}^{-1})^{-1}$. Clearly $T_{\rm{eff}}$ responds at all temperatures of the cryostat. Below $\sim30~$mK the effective temperature is $T_{\rm{eff}}=90~$mK. Above $\sim80~$mK, $T_{\rm{eff}}$ increases at the same rate as the cryostat temperature, indicating that the chip temperature is now limited by the phonon bath of the mixing chamber. The data in Fig.~\ref{figs7}(b) support the presence of an effective bath temperature of $\sim90~$mK when the cryostat is at the base temperature of $T_{B}=10~$mK, as was also inferred in Fig.\ref{figs7}(a) from the measurements of qubit spectra at different splittings. Other experiments with superconducting qubits have inferred similar effective temperatures \cite{fink_q-to-c}. $T_{\rm{eff}}$ is therefore a good indication of the effective system temperature and supports the observed changes in transmission for decreasing qubit splittings in Fig.~3 of the main text as having the origin in thermal effects and not dephasing. 

\begin{figure}[!hbt]
\centering
\includegraphics{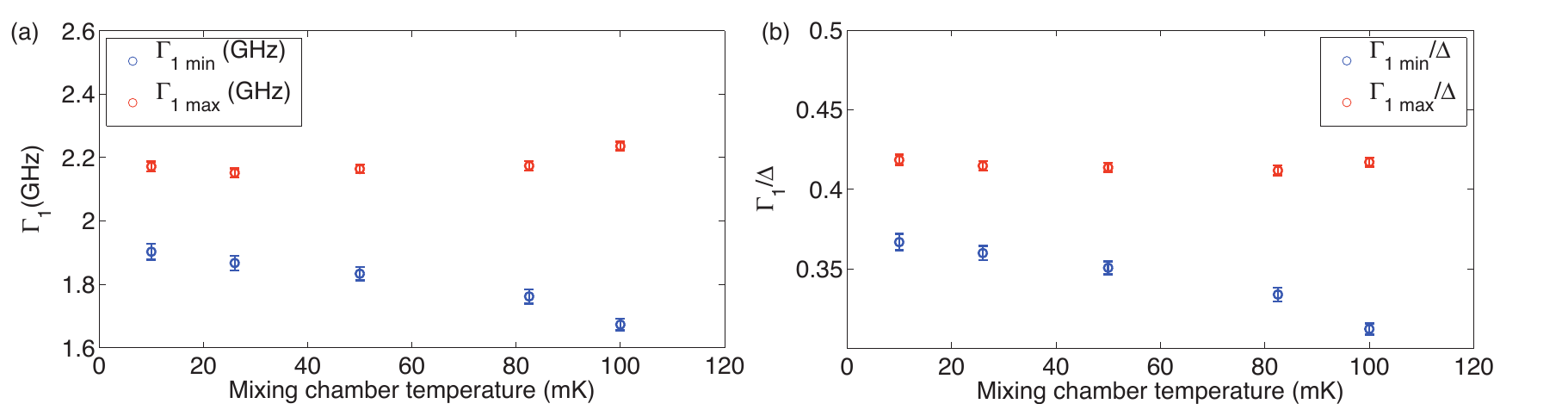}
\caption{\label{figs8}(a) Calculated bounds on qubit emission rate $\Gamma_1$ as function of temperature of cryostat. The average emission rate remains constant for the temperatures used. (b) Calculated bounds on normalized coupling $\Gamma_1/\Delta$ of qubit to transmission line. The maximum coupling rate remains constant approximately while the average decreases.}
\end{figure}

We can also calculate the bounds on the qubit emission rate $2\Gamma_2r_0<\Gamma_1<2\Gamma_2\sqrt{r_0}$, shown in Fig.~\ref{figs8}(a), and the normalized coupling $\Gamma_1/\Delta$ in Fig.~\ref{figs8}(b). The average emission rate $\Gamma_1$ remains constant up to 100~mK, while the average normalized coupling decreases slightly for increasing temperatures.

\section*{S6: Relation between $\Gamma_1$ and $\alpha_{\rm{SB}}$}

Let us begin with the spin-boson Hamiltonian 
\begin{equation}\label{eq:spinboson}
H= H_0+H_{\text{int}}=\frac{\hbar\Delta}{2}\sigma_z+\sum_k  \hbar\omega_ka^\dagger_k a_k +\sigma_x\sum_k g_k(a^\dagger_k+a_k),
\end{equation}
which is characterized by the spectral function, defined as
\begin{equation}
J(\omega)=\frac{2\pi}{\hbar^2}\sum_k g_k^2\delta(\omega-\omega_k)=\pi\omega\alpha_{\rm{SB}},
\label{eq:spinboson2}
\end{equation}
where we have assumed an Ohmic spectral bath. As it is usual in condensed matter physics \cite{spin-boson_rmp}, \cite{Weiss}, \cite{kondo}, we have expressed the spectral function $J(\omega)$ as function of a dimensionless constant $\alpha_{\rm{SB}}$, which characterizes the different quantum phases of the spin-boson model. More precisely, for $\alpha_{\rm{SB}}<1/2$ the system is in the Markovian regime, for $1/2<\alpha_{\rm{SB}}<1$ the system is in the overdamped regime, and for $\alpha_{\rm{SB}}>1$ the system is in the localized phase. Note that our definition of $J(\omega)$ differs from the one in \cite{spin-boson_rmp} due to a factor of $1/2$ that we omit in the last term of Eq.~(\ref{eq:spinboson}).

Our aim in this section is to relate the qubit decay rate $\Gamma_1$, obtained from the master equation formalism, to the parameter $\alpha_{\rm{SB}}$.

To this end, we will derive a quantum master equation for the qubit.
We start from the combined qubit-bath density matrix in the interaction picture
\begin{equation}
\rho(t)=U(t)\rho_0 U^\dagger(t),
\label{eq:dm}
\end{equation}
where the unitary transformation $U(t)=\exp(iH_0 t)$ brings us into the rotating frame. This yields the following time-evolution equation for the density matrix $\rho(t)$
\begin{equation}
\dot{\rho}=-\frac{i}{\hbar}[H_{\text{int}},\rho(t)],
\end{equation}
being $H_\text{int}(t)$ the coupling Hamiltonian in the interaction picture, given by
\begin{eqnarray}
H_{\text{int}}(t)&=&U(t)H_{\text{int}} U(t)^\dagger\nonumber\\
&=&(\sigma_{+}e^{i\Delta t}+\sigma_{-}e^{-i\Delta t})\sum_k g_k (a_k^\dagger e^{i\omega_k t}+a_k e^{-i\omega_k t})\nonumber\\
&=& A(t)X(t),
\end{eqnarray}
where $A(t)$, $X(t)$ are the system and bath operators, respectively.

Equation (\ref{eq:dm}) can be formally integrated, yielding the following integro-differential equation
\begin{equation}
\dot{\rho}(t)=\rho(0)-\frac{1}{\hbar^2}\int_0^t \text{d}\tau[H_\text{int}(t),[H_\text{int}(\tau),\rho_(\tau)]].
\end{equation}
As it is commonplace, we assume the Born approximation (weak coupling to the bath, which allows us to approximate $\rho(t)=\rho_\text{sys}(t)\otimes\rho_{\text{b}}(0)$, for any time $t$) and the Markov approximation (delta-correlated bath), which in turn corresponds with the Markovian dynamics of the spin-boson model defined by $\alpha_{\rm{SB}}<1/2$ \cite{juanjo_polaron}. Under these conditions, we find a second-order differential equation for the reduced density matrix of the system
\begin{equation}
\dot{\rho}_\text{sys}=-\frac{1}{\hbar^2}\int_0^t \text{d}\tau \text{Tr}_{\text{b}}[H_\text{int}(t),[H_\text{int}(\tau),\rho_\text{sys}(\tau)\otimes\rho_\text{b}(0)]], 
\end{equation}
where $\text{Tr}_\text{b}(A(t)X(t))$ refers to the trace over the bath degrees of freedom $X(t)$. 
Expanding the double commutator, and using the cyclic property of the trace, $\text{Tr}(AX)=\text{Tr} (XA)$, we can rewrite the master equation as
\begin{equation}
\dot{\rho}_\text{sys}(t)=\frac{\Gamma_1}{2}(2\sigma_{-}\rho_\text{sys}(t)\sigma_{+}-\sigma_{+}\sigma_{-} \rho_\text{sys}(t)-\rho_\text{sys}(t)\sigma_{+}\sigma_{-}),
\end{equation}
where the spontaneous decay rate $\Gamma_1$ is given by
\begin{eqnarray}
\Gamma_1&=&\frac{1}{\hbar^2}\int_{-\infty}^\infty \text{d}\tau e^{-i\Delta\tau}\left\langle[X(\tau),X(0)]_{+}\right\rangle\\
&=&\frac{1}{\hbar^2}\int_{-\infty}^\infty \text{d}\tau e^{-i\Delta\tau}\sum_kg_k^2[(1+n_k)e^{i\omega_k\tau}+n_ke^{-i\omega_k\tau}].\nonumber
\label{eq:gamma}
\end{eqnarray}
In equation (\ref{eq:gamma}), we have introduced the symetrized bath correlation function
\begin{eqnarray}
\left\langle[X(\tau),X(0)]_{+}\right\rangle&=&\text{Tr}_{\text{b}}[(X(\tau)X(0)+X(0)X(\tau))\rho_\text{b}(0)]\nonumber\\
&=&\sum_kg_k^2[(1+n_k)e^{i\omega_k\tau}+n_ke^{-i\omega_k\tau}], 
\end{eqnarray}
which can be readily calculated using the bosonic commutation relations $[a_k,a_{k'}]=0$, $[a_{k},a^\dagger_{k'}]=\delta_{k{k'}}$ and the 
two-time correlation functions
\begin{eqnarray}
\left\langle a^\dagger (t) a(t')\right\rangle&=&\sum_k g_k^2 n_k e^{i\omega_k(t-t')},\nonumber\\
\left\langle a (t) a^\dagger(t')\right\rangle&=&\sum_k g_k^2 (1+n_k) e^{i\omega_k(t'-t)}.
\end{eqnarray}
In the above expressions, $n_k$ is the average number of photons in the $k$-th oscillator, and is given by
\begin{equation}
n_k=\frac{1}{\exp(\hbar\omega_k/k_B T)-1}.
\end{equation}
For the sake of simplicity, but without loss of generality, we will assume that we are at zero temperature, so that $n_k=0$. Therefore, the relaxation rate $\Gamma_1$ can be rewritten as
\begin{eqnarray}
\Gamma_1&=&\frac{1}{\hbar^2}\int_{-\infty}^\infty \text{d}\tau e^{-i\Delta\tau}\sum_kg_k^2e^{i\omega_k\tau}\nonumber\\
&=&\frac{1}{\hbar^2}\sum_kg_k^2\int_{-\infty}^\infty \text{d}\tau e^{i(\omega_k-\Delta)\tau}.
\label{eq:gamma_2}
\end{eqnarray}
The last term in Eq.~(\ref{eq:gamma_2}) is nothing but the Fourier transform of the
delta function
\begin{equation}
\delta{(\omega_k)}=\frac{1}{2\pi}\int^{\infty}_{-\infty}\text{d}\tau e^{i\omega_k\tau},
\end{equation}
yielding the following expression for $\Gamma_1$
\begin{equation}
\Gamma_1=\frac{2\pi}{\hbar^2}\sum_kg_k^2\delta(\Delta-\omega_k)=J(\Delta).
\label{eq:gamma_3}
\end{equation}
Using the second identity of Eq.~(\ref{eq:spinboson2}) we finally arrive at a relation between $\Gamma_1$ and $\alpha_{\rm{SB}}$, 
\begin{equation}\label{eq:gamma-alpha}
\Gamma_1=\pi\alpha_{\rm{SB}} \Delta.
\end{equation}
It is worth mentioning that this result can be generalized for a bath at finite temperature $T$ \cite{Weiss}, \cite{gardiner}, \cite{schon_TLS}.  

\begin{center}
\begin{figure}[!hbt]
\includegraphics{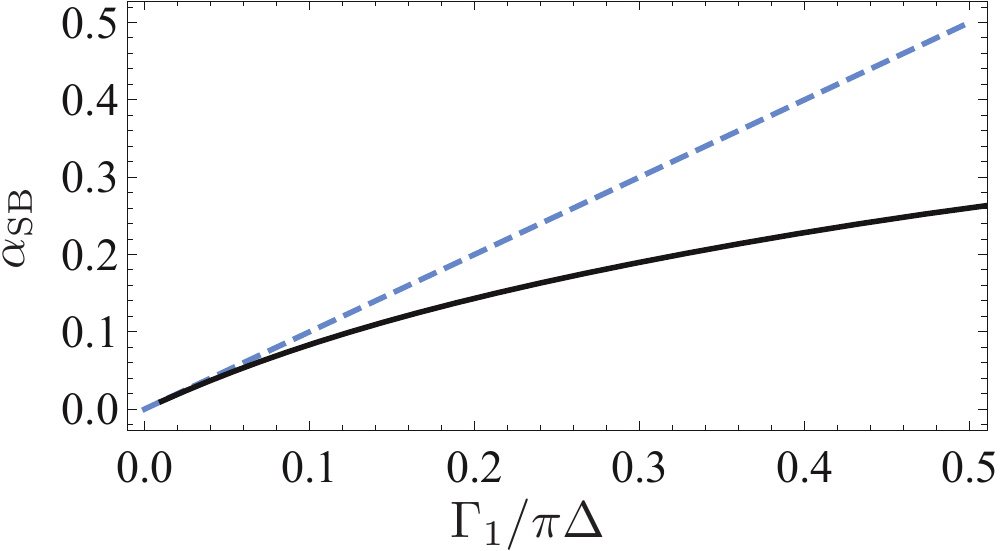}
\caption{\label{fig:gvsa} Polaron ansatz \cite{juanjo_prep} calculations of $\Gamma_1/\Delta$ as function of $\alpha_{\rm{BS}}$ (solid-black line) compared to the Born-Markov approximation (dashed-blue line), equation~(\ref{eq:gamma-alpha}). Clearly, equation~(\ref{eq:gamma-alpha}) is a lower bound for $\alpha_{\rm{SB}}>0.1$.}
\end{figure}
\end{center}

Eq.~(\ref{eq:gamma-alpha}) is valid in the Born-Markov and rotating-wave approximations. It is known from the spin-boson model that up to $\alpha_{\rm{SB}}=1/2$ (see equation~5.23 from reference~\cite{spin-boson_rmp}), corresponding to $\Gamma_1/\Delta\sim1$ and therefore well within the ultrastrong coupling regime, Eq.~(\ref{eq:gamma-alpha}) is still correct. The regime $0.5<\alpha_{\rm{SB}}<1$ presents more difficulties, as the spin-boson model becomes nonperturbative. Using a polaron transformation \cite{juanjo_polaron}, an analytical model has been found \cite{juanjo_prep} to yield correct results for $\alpha_{\rm{SB}}>0.1$. Using this technique we calculate values for $\Gamma_1/\Delta$ as function of $\alpha_{\rm{SB}}$ and compare it to equation~(\ref{eq:gamma-alpha}), shown in Fig.~\ref{fig:gvsa}. The results show that equation~(\ref{eq:gamma-alpha}) is a lower bound for $\alpha_{\rm{SB}}>0.1$. We assume in the analysis of our results for $\Gamma_1/\Delta>1.5$ that equation~(\ref{eq:gamma-alpha}) remains a lower bound.

\section*{S7: The maximum coupling}
As detailed in \cite{borja_usc}, a flux qubit coupled to a transmission line, sharing a junction, can be calculated from the case of coupling to a single-mode resonator. We derive here the expression of the coupling rate that is used in Fig.~4(a) of the main text to fit the experimental normalized coupling rate $\Gamma_1/\Delta$. 

The quantized flux field in a 1D-transmission line assuming periodic boundary conditions (suitable for infinite transmission lines) takes the form
\be
\hat{\phi} = \sum_k\sqrt{\frac{\hbar}{2c_0\omega_kL}}\left(\hat{a}_k e^{i(kx - \omega_kt)}+\hat{a}_k^{\dag} e^{-i(kx - \omega_kt)}\right),
\ee
where the line has length $L$, capacitance and inductance per unit length $c_0, l_0$ and mode frequency $\omega_k$. The dispersion relation is given by $\omega_k = kc = k(l_0c_0)^{-1/2}$, $c$ being the speed of light in the line. The coupling term takes the form (see Supplementary material in \cite{borja_usc}):
\be
\hat{H}_{\mathrm{int}} =\varphi_0\hat{\varphi}_{\beta}\frac{1}{l_0} \frac{\partial\hat{\phi}}{\partial x}\delta(x),
\ee
which is nothing but the current along the transmission line times the effective node flux generated by the qubit $\varphi_0\hat{\varphi}_{\beta}$, with $\varphi_0 = \Phi_0/2\pi$ the reduced flux quantum and $\hat{\varphi}_{\beta}$ the phase operator across the qubit coupling junction $\beta$. $\delta(x)$ is the Dirac delta, since the qubit is assumed to sit at the origin $x=0$. The strength of the coupling to mode $k$ is given by \cite{borja_usc} 
\be
g_k = \frac{1}{l_0}\varphi_0\varphi_{\beta}\frac{1}{\sqrt{L}}\sqrt{\frac{\hbar\omega_k}{2c_0c^2}},
\ee
with $\varphi_{\beta}\equiv\langle1|\hat{\varphi}_{\beta}|0\rangle$ the matrix element of the phase operator across the qubit coupling junction $\beta$. The spectral density $J(\omega)$ \cite{caldeira-legget}, which as shown in section S6 corresponds to the spontaneous emission rate $\Gamma_1$, can be directly calculated
\be
J(\omega) = 2\pi\sum_k(|g_k|/\hbar)^2\delta(\omega-\omega_k)=2\pi\sum_k\frac{1}{l_0^2}\frac{1}{L}\frac{\omega_k}{2\hbar c_0c^2}\varphi_0^2|\varphi_{\beta}|^2\delta(\omega-\omega_k).
\ee
Taking the limit to the continuum, using that the density of states is $L/2\pi$,
\be
J(\omega) = 2\int_0^{\infty}\mathrm{d}\omega_k\frac{\omega_k}{2\hbar c_0l_0^2c^3}\varphi_0^2|\varphi_{\beta}|^2\delta(\omega-\omega_k) = \frac{\omega}{\hbar Z_0}\varphi_0^2|\varphi_{\beta}|^2,
\ee
$Z_0=(l_0/c_0)^{1/2}$ being the characteristic impedance of the transmission line. The factor of 2 in front of the integral is due to the fact that the frequency $\omega_k$ is degenerate for wavectors $k$ and $-|k|$. By  integrating over $k<0$ and $k>0$ we are taking into account the current fluctuations of the two semi-infinite transmission lines, which represent two independent baths. Therefore, and connecting to the traces in Fig.~4(a) of the main text, we can express the reduced coupling $\Gamma_1/\Delta$ as function of the expectation value of the phase operator and the impedance of the line:
\be\label{eqdeltagamma}
\frac{J(\Delta)}{\Delta}=\frac{\Gamma_1}{\Delta}=\frac{1}{4e^2}\frac{\hbar}{Z_0}|\varphi_{\beta}|^2 = \frac{1}{2\pi}\frac{R_Q}{Z_0}|\varphi_{\beta}|^2,
\ee
where $R_Q = h/(2e)^2\simeq6.5~$k$\Omega$ is the resistance quantum. Equation~(\ref{eqdeltagamma}) indicates that in order to increase the coupling to its highest value, $Z_0$ has to be as low as possible and $|\varphi_{\beta}|$ must be increased by making the $\beta$-junction size smaller and therefore having a phase drop of order 1 across it. Achieving $\Gamma_1/\Delta\approx10$ is therefore within reach. From this analysis the quantity $\Gamma_1/\Delta$ can be understood as a normalized coupling strength. 

Equation~(\ref{eqdeltagamma}) has the same validity as equation~(\ref{eq:gamma-alpha}) since it relies on equation~(\ref{eq:gamma_3}). Therefore it is a lower bound for the range $0.5<\alpha_{\rm{SB}}<1$, or $1.5<\Gamma_1/\Delta<3$. This is verified in our experiment where in Fig.~4(a) the values of $\Gamma_1/\Delta$ lie above the curves for $\beta<2$, where $\alpha_{\rm{SB}}>0.5$.

\end{document}